\documentclass[a4paper,fleqn,usenatbib]{mnras}
\usepackage{amsmath}
\usepackage{mathptmx}
\usepackage{txfonts}
\usepackage[T1]{fontenc}
\usepackage{ae,aecompl}
\usepackage{graphicx}	
\usepackage{amssymb}	

\newcommand{\e}{\end{equation}}
\newcommand{\bear}{\begin{eqnarray}}
\newcommand{\ear}{\end{eqnarray}}
\def\aj{AJ}
\def\apj{ApJ}
\def\apjs{ApJS}
\def\jcap{JCAP}
\def\mnras{MNRAS}
\def\prl{Physical Review Letters}
\def\aap{A\&A}
\def\prd{Physical Review D}
\def\nat{Nature}
\def\apjs{ApJS}
\def\apjl{ApJ Letters}

\title[Testing isotropy with Shannon entropy]{A new method for
  testing isotropy with Shannon entropy}
    
\author[Pandey, B.]  {Biswajit Pandey\thanks{E-mail:
    biswap@visva-bharati.ac.in}\\Department of Physics, Visva-Bharati
  University, Santiniketan, Birbhum, 731235, India\\ 
}
       
 \date{\today}

 \pubyear{2015}
  
\begin{document}
\label{firstpage}
\pagerange{\pageref{firstpage}--\pageref{lastpage}}      
\maketitle
       
 \begin{abstract}

 We propose a method for testing isotropy of a three-dimensional
 distribution using Shannon entropy. We test the method on some Monte
 Carlo simulations of isotropic and anisotropic distributions and find
 that the method can effectively identify and characterize different
 types of hemispherical asymmetry inputted in a distribution. We
 generate anisotropic distributions by introducing pockets of
 different densities inside homogeneous and isotropic distributions
 and find that the proposed method can effectively quantify the degree
 of anisotropy and determine the geometry of the pockets
 introduced. We also consider spherically symmetric radially
 inhomogeneous distributions which are anisotropic at all points other
 than the centre and find that such anisotropy can be easily
 characterized by our method. We use a semi analytic galaxy catalogue
 from the Millennium simulation to study the anisotropies induced by
 the redshift space distortions and find that the method can separate
 such anisotropies from a general one. The method may be also suitably
 adapted for any two dimensional maps on the celestial sphere to study
 the hemispherical asymmetry in other cosmological observations.

 \end{abstract}

       \begin{keywords}
         methods: numerical - galaxies: statistics - cosmology: theory - large
         scale structure of the Universe.
       \end{keywords}
       
       \section{Introduction}

The Cosmological principle which assumes that the Universe is
statistically homogeneous and isotropic on sufficiently large scales
is one of the fundamental assumptions of modern cosmology. This
assumption can not be proved in a rigorously mathematical sense but
can be verified from various cosmological observations. Testing the
assumption of statistical homogeneity and isotropy is important as our
interpretations of various cosmological observations are based on our
current understanding of the Universe which in turn relies on the
cosmological principle. Homogeneity and isotropy are two different but
related aspects which may or may not coexist in a system. The Universe
can be homogeneous without being isotropic or can be isotropic around
a point without being homogeneous. But isotropy around each and every
point guarantees homogeneity.

Observationally the most powerful evidence for isotropy is provided by
the the near uniform temperature of the Cosmic Microwave Background
Radiation (CMBR) across the whole sky
\citep{penzias,smoot,fixsen}. However the CMBR is not completely
isotropic. Over the years many studies have reported power asymmetries
and unlikely alignments of low multipoles
\citep{schwarz1,land,hanlewis,moss,grupp,dai}. The asymmetries found
in WMAP were largely attributed to deficiencies in the foreground
subtraction \citep{bennett} and non circularity of beams
\citep{das}. However recent analysis of PLANCK data by
\citet{adeplanck1} reported that the power asymmetry persist on scales
corresponding to $l\sim 600$ and the deviations from isotropy are at
high statistical significance ($3-\sigma$). A further analysis using
multi frequency PLANCK data by \citet{adeplanck2} confirmed the power
asymmetry and show that the foreground residuals are unlikely to
affect these results. A multitude of other evidences favouring
isotropy comes from isotropy in the angular distributions of radio
sources \citep{wilson,blake}, isotropy in the X-ray background
\citep{wu,scharf}, isotropy of Gamma-ray bursts \citep{meegan,briggs},
isotropy in the distribution of galaxies \citep{marinoni,alonso},
isotropy in the distribution of supernovae \citep{gupta,lin} and
isotropy in the distribution of neutral hydrogen
\citep{hazra}. Although large number of studies favour the statistical
isotropy of the Universe on large scales there is no clear consensus
on this issue yet. There are some studies with Type-Ia supernovae
which find evidence for statistically significant anisotropy
\citep{schwarz2,campanelli,kalus,javanmardi,bengaly}. Some other
studies with radio sources \citep{jackson} and galaxy luminosity
function \citep{appleby} also point towards significant anisotropy.
These anisotropies may originate from the systematics in the data. But
there is also a possibility that they signal the failure of the
assumption of cosmic isotropy itself. There would be a major paradigm
shift in modern cosmology if the assumption of cosmic isotropy is
ruled out with high statistical significance by multiple data sets. A
large number of theoretical studies have been carried out on the
possible origins of such anisotropy and their consequences
\citep{chan,shtanov,barrow,soda,pitrou,marrozi,subho}.

Different statistics \citep{hajian,zunckel,taylor} has been developed
to test isotropy for different types of data sets. \citet{pandey}
introduce a method based on the Shannon entropy \citep{shannon48} for
characterizing inhomogeneities in a 3D distribution of points and
applied the method on some Monte Carlo simulations of inhomogeneous
distributions and N-body simulations which show that the proposed
method has great potential for testing the large scale homogeneity in
galaxy redshift surveys. Recently \citet{pandey15} applied this method
to galaxy distributions from SDSS DR12 \citep{alam} and find that the
inhomogeneities in the galaxy distributions persist at least upto a
length scale of $120 \, h^{-1}\, {\rm Mpc}$. A subsequent analysis of
the SDSS LRG distribution \citep{eisen} by \citet{pandey16} using an
improved method reveal that the LRG distribution is homogeneous beyond
length scales of $\sim 150 \, h^{-1}\, {\rm Mpc}$. In the present work
we propose a method for testing isotropy based on the Shannon
entropy. The proposed method can be directly applied to 3D
distributions of galaxies or the corresponding density fields derived
from them. The method can be also easily extended to 2D maps such as
CMB and may be also suitably adapted for testing isotropy of
gravitational waves, cosmic rays, X-ray or Radio sky.

A brief outline of the paper follows. We describe our method in
Section 2, describe the data and the tests in Section 3 and present
the results in Section 4 and Conclusions in Section 5.

We have used a $\Lambda$CDM cosmological model with $\Omega_{m0}=0.3$,
$\Omega_{\Lambda0}=0.7$ and $h=1$ throughout.


\section{METHOD OF ANALYSIS}

\citet{pandey} propose a method based on the Shannon entropy to study
inhomogeneities in a 3D distribution of points. Shannon entropy
\citep{shannon48} is originally proposed by Claude Shannon to quantify
the information loss while transmitting a message in a communication
channel. It gives a measure of the amount of information required to
describe a random variable. The Shannon entropy for a discrete random
variable x with $n$ outcomes $\{x_{i}:i=1,....n\}$ is a measure of
uncertainty denoted by $H(X)$ defined as,

\begin{equation}
H(X) =  - \sum^{n}_{i=1} \, p(x_{i}) \, \log \, p(x_{i})
\label{eq:shannon1}
\end{equation}

where $p(x)$ is the probability distribution of the random variable
x.  

In the present work we propose information entropy as a measure of
isotropy. We assume a set of points distributed in 3D and wish to test
the assumption of statistical isotropy around any point.  In order to
test the isotropy around a point we first uniformly bin $\cos\theta$
and $\phi$ where $\theta$ and $\phi$ are the polar angle and azimuthal
angle in spherical polar co-ordinates respectively. Uniform binning of
$\cos\theta$ and $\phi$ ensure equal size for each solid angle bin
since $d\Omega=\sin\theta d\theta d\phi$. The number of bins
$m_{\theta}$ and $m_{\phi}$ for binning $\cos\theta$ and $\phi$ are to
be decided conveniently. A given choice of $m_{\theta}$ and $m_{\phi}$
results in a total $m_{total}=m_{\theta}m_{\phi}$ solid angle bins. We
impose an upper limit to the radius $r$ upto which the isotropy is to
be tested. There is a natural limit to $r$ from the fact that the data
points are available only upto a certain radius $r_{max}$.

We pick up a point about which isotropy has to be tested and treating
that point as origin define the coordinates of all the other points in
the distribution.  We bin the co-ordinates for a given choice of
$m_{\theta}$, $m_{\phi}$ and $r$. This results in
$m_{total}=m_{\theta}m_{\phi}$ volume elements each covering the same
solid angle. For any given value of $r$ each of the volume elements
has the same radial extension ensuring same volume
$dv=\frac{r^3}{3}d\Omega$ for each of them. We count the number of
points $n_i$ inside each of the $m_{total}$ volume elements where $i$
is the index of the volume element. In general each galaxy within
radius $r$ from the centre can reside in only one of the $m_{total}$
volume elements. But which volume element a particular galaxy belongs
to ?  The answer to this question has $m_{total}$ likely outcomes. We
define a random variable $X_{\theta\phi}$ which has $m_{total}$
possible outcomes each given by,
$f_{i}=\frac{n_{i}}{\sum^{m_{total}}_{i=1} \, n_{i}}$ with the
constraint $\sum^{m_{total}}_{i=1} \, f_{i}=1$.  The Shannon entropy
associated with the random variable $X_{\theta\phi}$ can be written
as,
\begin{eqnarray}
H_{\theta\phi}(r)& = &- \sum^{m_{total}}_{i=1} \, f_{i}\, \log\, f_{i} \nonumber\\ &=& 
\log N - \frac {\sum^{m_{total}}_{i=1} \, n_i \, \log n_i}{N}
\label{eq:shannon2}
\end{eqnarray}
Where $N$ is the total number of points within radius $r$. The base of
the logarithm is arbitrary and we choose it to be $10$.  $f_{i}$ will
have the same value $\frac{1}{m_{total}}$ for all the volume elements
when $n_{i}$ is same for all of them. This maximizes the Shannon
entropy to $(H_{\theta\phi})_{max}=\log \, m_{total}$ for a given
choice of $m_{\theta}$,$m_{\phi}$ and any $r$. We define the relative
Shannon entropy as the ratio of the entropy of a random variable
$X_{\theta\phi}$ to the maximum possible entropy
$(H_{\theta\phi})_{max}$ associated with it. The relative Shannon
entropy $\frac{H_{\theta\phi}(r)}{(H_{\theta\phi})_{max}}$ then
quantifies the degree of uncertainty in the knowledge of the random
variable $X_{\theta\phi}$. Equivalently
$a_{\theta\phi}(r)=1-\frac{H_{\theta\phi}(r)}{(H_{\theta\phi})_{max}}$
quantify the residual information and can be treated as a measure of
anisotropy. The fact that galaxies are not residing in any particular
volume element and rather are distributed across all of them with
different probabilities acts as a source of information. If all of
them would have been residing in a particular volume element then
there would be no uncertainty and no information at all making
$H_{\theta\phi}=0$ or $a_{\theta\phi}=1$. This fully determined
situation corresponds to maximum anisotropy. On the other hand when
all the $m_{total}$ volume elements are populated with equal
probabilities it would be most uncertain to decide which particular
volume element a galaxy belongs to. This maximizes the information
entropy to $H_{\theta\phi}=\log \, m_{total}$ turning
$a_{\theta\phi}=0$. This corresponds to a situation when the
distribution is completely isotropic. The galaxy distribution is
expected to be anisotropic on small scales but with increasing solid
angle $d\Omega$ and radius $r$ one would expect it to be isotropic on
some scale provided the Cosmological principle holds on large
scales. We change the value of $r$ starting from a small radius
$r$ and gradually increase it in steps upto the maximum radius
$r_{max}$ to study how $a_{\theta\phi}(r)$ varies with $r$ for a given
choice of $m_{\theta}$ and $m_{\phi}$. It may be noted here that the
analysis can be also done for data covering parts of the sky.

We would like to mention here that the proposed measure of anisotropy
would never be exactly zero and would be also sensitive to binning and
sub-sampling. So we adopt a workable definition of isotropy where the
distribution is considered to be isotropic when the measured
anisotropy lies within the $1-\sigma$ errorbars of the anisotropy
expected for a Poisson distribution. Consequently our preferred
binning and sampling would be such that for which the anisotropy in
the Poisson distribution decays to approximately zero within the
scales probed.

Besides the radial anisotropy one can also measure the degree of polar
anisotropy $a_{\phi}(\theta)=1-\frac{H_{\phi}}{(H_{\phi})_{max}}$ and
the azimuthal anisotropy
$a_{\theta}(\phi)=1-\frac{H_{\theta}}{(H_{\theta})_{max}}$ as function
of $\theta$ and $\phi$ by carrying out the sum respectively over
$m_{\phi}$ or $m_{\theta}$ instead of $m_{total}$ in
\autoref{eq:shannon2}. Note that in this case $N$ would be the total
number of points inside all the $m_{\phi}$ or $m_{\theta}$ volume
elements at different $\theta$ or $\phi$ respectively. Note that
$a_{\phi}(\theta)$ measures isotropy among all the $\phi$ bins at each
$\theta$ and similarly $a_{\theta}(\phi)$ measures isotropy among all
the $\theta$ bins at each $\phi$. $a_{\phi}(\theta)$ and
$a_{\theta}(\phi)$ are then determined at different $\theta$ and
$\phi$ values respectively. One can also study $a_{\phi}(r)$ and
$a_{\theta}(r)$ as a function of $r$ at fixed $\theta$ and $\phi$
values respectively. However in the present work we only employ
$a_{\theta\phi}(r)$, $a_{\phi}(\theta)$ and $a_{\theta}(\phi)$ to
quantify and characterize the anisotropies present in a distribution.

It would be worth mentioning here that one can also estimate the
variance in the number counts across the various volume elements in
different ($\theta$,$\phi$) directions to measure the anisotropy
present in a distribution. But we prefer entropy because unlike
variance it is related to the higher order moments of a
distribution. So in principle the entropy can be a better measure of
non-uniformity than variance as it uses more information about the
probability distribution. The Variance and the entropy would be
equivalent as a measure of non-uniformity only when the probability
distribution is fully characterized by the first two moments such as
in a Gaussian distribution. However even all the higher order moments
together can not uniquely describe a highly tailed distribution in the
non-linear regime \citep{carron1,carron3}. It has been suggested that
the spectrum of the log-density field carries more information than
the spectrum of the field and could be a better choice in such cases
\citep{carron2}.

When the probability distribution function is Gaussian then the
associated Shannon entropy is $\log \sqrt{2 \pi e} \sigma$ where
$\sigma$ is the standard deviation of the distribution. In $\Lambda
CDM$ model one can use the variance of the smoothed density field
predicted from the power spectrum to estimate the Shannon entropy
associated with that scale provided the density field is assumed to be
Gaussian.

One can integrate $-\int p(x)\,\log p(x)\,dx$ for any probability
distribution to estimate the associated differential entropy of the
corresponding distribution. For example the differential entropy of
the Poisson distribution is given by, $H(p_{\lambda})=\lambda
\log\frac{e}{\lambda}+e^{-\lambda}\sum_{k=0}^{\infty}\frac{\lambda^{k}\log
  k!}{k!}$, where $\lambda$ is the average number of events per
interval. In this case $\lambda$ is the average number of points per
volume element. Although the discrete entropy and differential entropy
have similar mathematical forms there are some important differences
between them. The differential entropy is not a number as in discrete
entropy, but rather a function of one or more parameters that
describes the associated probability distribution. The differential
entropy does not provide the average amount of information contained
in a random variable like its discrete counterpart. It is not an
absolute measure of uncertainty rather it measures relative
uncertainty or changes in uncertainty. One can show that the
discrete entropy $H(x)_{\Delta}$ is related to the differential
entropy $H(x)$ as, $H(x)_{\Delta} \approx H(x)-\log \Delta$ in the
limit $\Delta \longrightarrow 0$ where $\Delta$ is the bin size used
in discretization. The extra term $-\log \Delta$ approaches infinity
as $\Delta \longrightarrow 0$. So one can only make relative
comparisons of differential entropies and a simple comparison between
the discrete entropy and the differential entropy is not quite
meaningful \citep{diff}.

The method presented here can be directly applied to different galaxy
redshift surveys to test the isotropy of the galaxy distributions in
the present Universe. The redshift surveys map the mass distribution
on a light cone time slice where the distribution does not evolve much
over the light crossing time of the survey. But for very large galaxy
surveys possible evolutionary effects can introduce signatures of
anisotropy in the data. Redshift dependent selection effects can also
introduce artificial anisotropy in the data. Besides these the
redshift space distortions is one of the most important source of
anisotropy in galaxy surveys. On large scales structures are
compressed along the line of sight due to coherent flows into
overdense regions and out of underdense regions whereas on small
scales structures are elongated along the line of sight by random
motions in virialized clusters.  The volume elements used for
measurement of $n_{i}$ in our method radially extends along the line
of sight where the radial extension is much larger compared to their
angular width. When measurements are done from the point from which
observations are carried out one would expect uniformity in the
measurement of $n_{i}$ across all directions for large $r$ provided
the Universe is isotropic. But this would appear anisotropic if one
shifts the origin from the point of observation given such sources of
anisotropies are present. It is important to distinguish the presence
of genuine anisotropies from the artificial ones such as introduced by
radial inhomogeneities (due to selection effects, evolutionary
effects) and redshift space distortions. We will show that our method
can distinguish the signatures of different kind of anisotropies
present in the distribution.

The method presented here has a significant advantage compared to the
method proposed by \citet{pandey} for testing homogeneity using
Shannon entropy. The volume elements used for measuring the number
count $n_{i}$ in this method do not overlap. Consequently all the
complexities due to overlap can be bypassed allowing one to have a
more direct and clear interpretation.

\section{TESTING THE METHOD}

In order to study the prospects and limitations of the proposed method
we carry out some preliminary tests by applying it to some simple
distributions. We consider the following distributions: (1)
homogeneous and isotropic Poisson distributions, (2) anisotropic
distributions generated by inserting pockets of different densities at
different locations in homogeneous and isotropic Poisson
distributions, (3) radially inhomogeneous Poisson distributions which
are isotropic only about one point i.e. the centre and (4) simulated
galaxy distributions from N-body simulations in real space and
redshift space.  

For the first three types we generate a set of Monte Carlo
realizations. The distributions of type (1) are isotropic and type (2)
are anisotropic by construction. The distributions of type (3) are
radially inhomogeneous and the radial variations are identical in all
directions making them isotropic about the centre of the sphere. But
if we shift the origin from the centre of the sphere the distribution
would appear anisotropic and the degree of anisotropy would depend on
the magnitude and direction of the shift in a predictable manner. For
the distributions of type (4) we use the data from a semi analytic
galaxy catalogue from the Millennium simulation. An isotropic
distribution in real space would appear anisotropic in redshift space
due to redshift space distortions induced by the peculiar velocities.
We map the particles in N-body simulations from real space to redshift
space using their peculiar velocities and measure the resulting
anisotropies induced by redshift space distortions. In all cases
we have considered a spherical region of radius $200 h^{-1} \, {\rm
  Mpc}$.

We analyze the datasets separately using the method described in
section 2. We divide the $\theta-\phi$ space into $m_{\theta}m_{\phi}$
solid angle bins where $m_{\theta}$ and $m_{\phi}$ are variables and
chosen conveniently. The following calculations are carried out for
each of the datasets described in the subsections below.

(i) We choose the minimum and maximum values of radius $r$ to be
$r_{min}=5 h^{-1} \, {\rm Mpc}$ and $r_{max}=200 h^{-1} \, {\rm
  Mpc}$. We gradually increase the radius $r$ in steps of $5 h^{-1} \,
{\rm Mpc}$ from $r_{min}$ to $r_{max}$ and compute the Shannon entropy
$\frac{H_{\theta\phi}}{(H_{\theta\phi})_{max}}$ for each radius using
all the available $m_{\theta}m_{\phi}$ bins.

(ii) We fix the radius at $r_{max}=200 h^{-1} \, {\rm Mpc}$ and compute
$\frac{H_{\theta}}{(H_{\theta})_{max}}$ for each $\theta$ using all the
  $m_{\phi}$ azimuthal bins available.

(iii) We fix the radius at $200 h^{-1} \, {\rm Mpc}$ and compute
  $\frac{H_{\phi}}{(H_{\phi})_{max}}$ for each $\phi$ using all the
 $m_{\theta}$ polar bins available.

(iv) We shift the origin by $100 h^{-1} \, {\rm Mpc}$ along the x-axes
without any rotation and repeat (i),(ii) and (iii) using $r_{max}=100
h^{-1} \, {\rm Mpc}$.

(v) We shift the origin by $100 h^{-1} \, {\rm Mpc}$ along the y-axes
without any rotation and repeat (i),(ii) and (iii) using $r_{max}=100
h^{-1} \, {\rm Mpc}$.

(vi) We shift the origin by $100 h^{-1} \, {\rm Mpc}$ along the z-axes
without any rotation and repeat (i),(ii) and (iii) using $r_{max}=100
h^{-1} \, {\rm Mpc}$.
 
In general one can apply the shift along any arbitrary directions and
it would not make any difference given the distribution is
isotropic. Further if one can verify the isotropy around other points
it would also help us in testing homogeneity of the distribution.

\subsection{MONTE CARLO SIMULATIONS}

The Monte Carlo simulations for the data sets of type (1) and (3) are
generated by considering two simple radial density distributions
$\rho(r,\theta,\phi) = K \, \lambda(r)$ where $\lambda(r) = 1$ for
type (1) and $\lambda(r) = \frac{1}{r^{2}}$ for type (3)
distributions. Here $K$ is a normalization constant. The type (1)
distributions are homogeneous and isotropic Poisson point process
which has a constant density everywhere.  The type (3) distributions
are radially inhomogeneous Poisson distributions which are isotropic
only about the centre.

Enforcing the desired number of particles $N$ within radius $R$ one
can turn the radial density function into a probability function within
$r=0$ to $r=R$ which is normalized to one when integrated over that
interval. So the probability of finding a particle at a given radius
$r$ is $P(r)=\frac{ r^2 \lambda(r) } {\int_{0}^{R} r^2 \lambda(r) \,
  dr}$ which is proportional to the density at that radius implying
more particles in high density regions.

We generate the Monte Carlo realizations of these distributions using
a Monte Carlo dartboard technique. The maxima of the function $r^2
\lambda(r)$ in $P(r)$ is at $r=R$ for type (1) distribution whereas in
type (3) distribution it is same and constant everywhere. We label the
maximum value of $P(r)$ as $P_{max}$. We randomly choose a radius $r$
in the range $0 \le r \le R$ and a probability value is randomly
chosen in the range $0 \le P(x) \le P_{max}$. The actual probability
of finding a particle at the selected radius is then calculated using
expression for $P(r)$ and compared to the randomly selected
probability value. If the random probability is less than the
calculated value, the radius is accepted and assigned isotropically
selected angular co-ordinates $\theta$ and $\phi$, otherwise the
radius is discarded. In this way, radii at which particle is more
likely to be found will be selected more often because the random
probability will be more frequently less than the calculated actual
probability. We choose $R=200 h^{-1} \, {\rm Mpc}$ and $N=10^{5}$.

To generate the distributions of type (2) we first generate a
homogeneous and isotropic Poisson distribution within a spherical
region of radius $200 h^{-1} \, {\rm Mpc}$ with $N=10^{5}$. We
randomly identify a region in $(r,\theta,\phi)$ space assuming the
centre of the sphere as origin. We discard all the data points from
the selected region and subsequently populate it with a homogeneous
and isotropic Poisson distribution having a different intensity
parameter than the original one. As a result this region will have a
different mean density which introduces a preferred direction and
hence anisotropy in the distribution.

We generate $10$ realizations for each of the above density
distributions and analyze them separately using the method described
earlier.

\subsection{MILLENNIUM SIMULATION}
Semi analytic models
\citep{white2,kauff1,kauff2,kauff3,cole1,cole2,somervil,bagh,benson,springel,guo}
provide a very powerful tool to study galaxy formation and
evolution. Galaxy formation and evolution involve many physical
processes such as gas cooling, star formation, supernovae feedback,
metal enrichment, merging and morphological evolution. The semi
analytic models parametrise the physics involved in terms of simple
models following the dark matter merger trees over time. The models
provide the statistical predictions of galaxy properties at some epoch
and the precision of these predictions are directly related to the
accuracy of the input physics. In the present work we use a
semi-analytic galaxy catalogue generated by \citet{guo} from the
Millennium Run simulation\citep{springel} who updated the previously
available galaxy formation models \citep{springel, croton, delucia}
with improved versions. The spectra and magnitude of the model
galaxies were computed using population synthesis models of
\citet{bruzual}. We place the origin at the centre of the simulation
box which has a length of $500 h^{-1} \, {\rm Mpc}$ on each side and
then identify all the galaxies within a radius of $200 h^{-1} \, {\rm
  Mpc}$ having r-band Petrosian absolute magnitude in the range
$-22\leq M_{r} \leq -20$. We randomly select $10^{5}$ galaxies from
them to construct the simulated galaxy sample in real space for our
analysis. We then map these galaxies to redshift space using their
peculiar velocities to obtain their distribution in redshift space.

\begin{figure*}
\resizebox{9cm}{!}{\rotatebox{-90}{\includegraphics{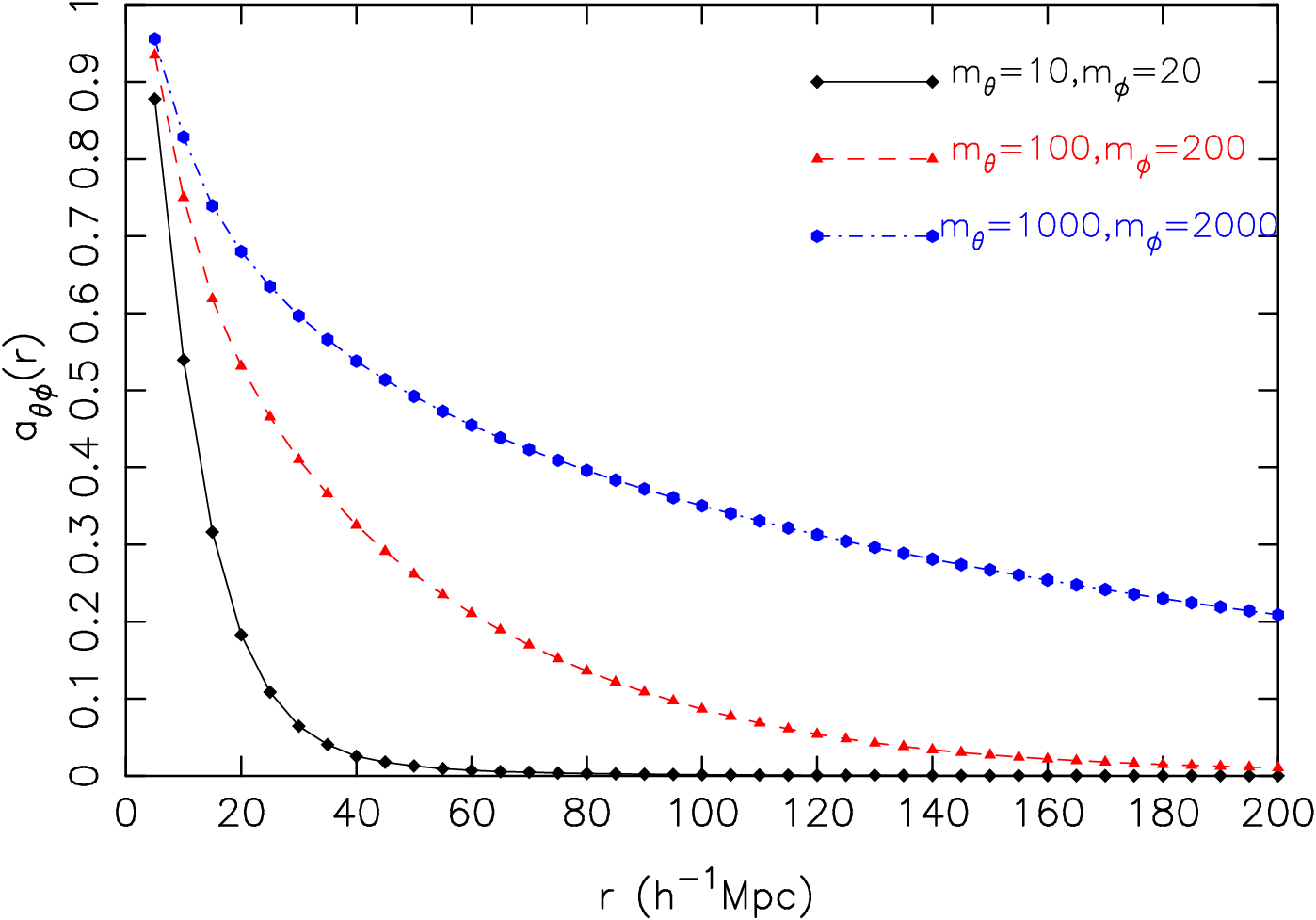}}}%
\resizebox{9cm}{!}{\rotatebox{-90}{\includegraphics{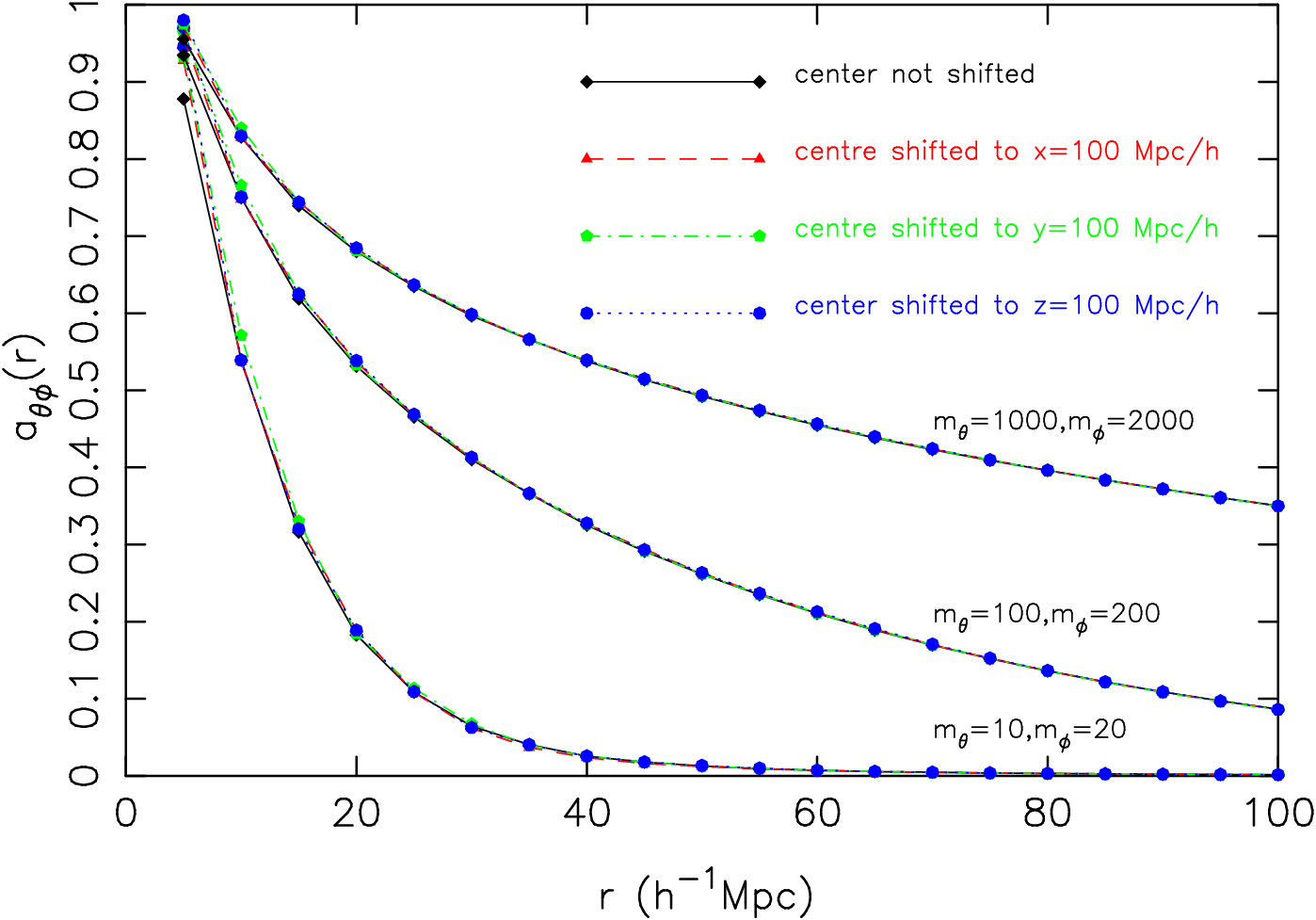}}}\\
\resizebox{9cm}{!}{\rotatebox{-90}{\includegraphics{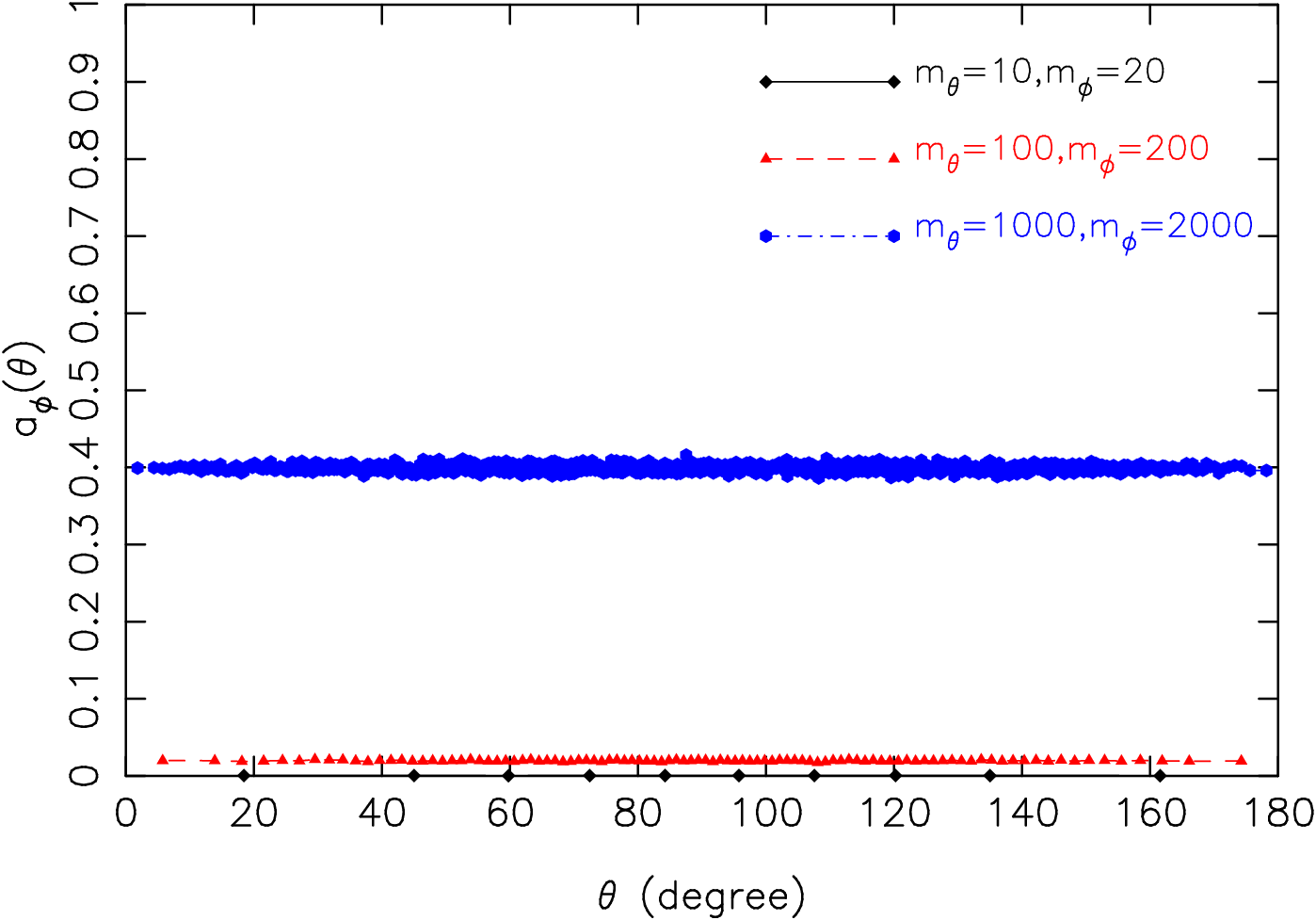}}}%
\resizebox{9cm}{!}{\rotatebox{-90}{\includegraphics{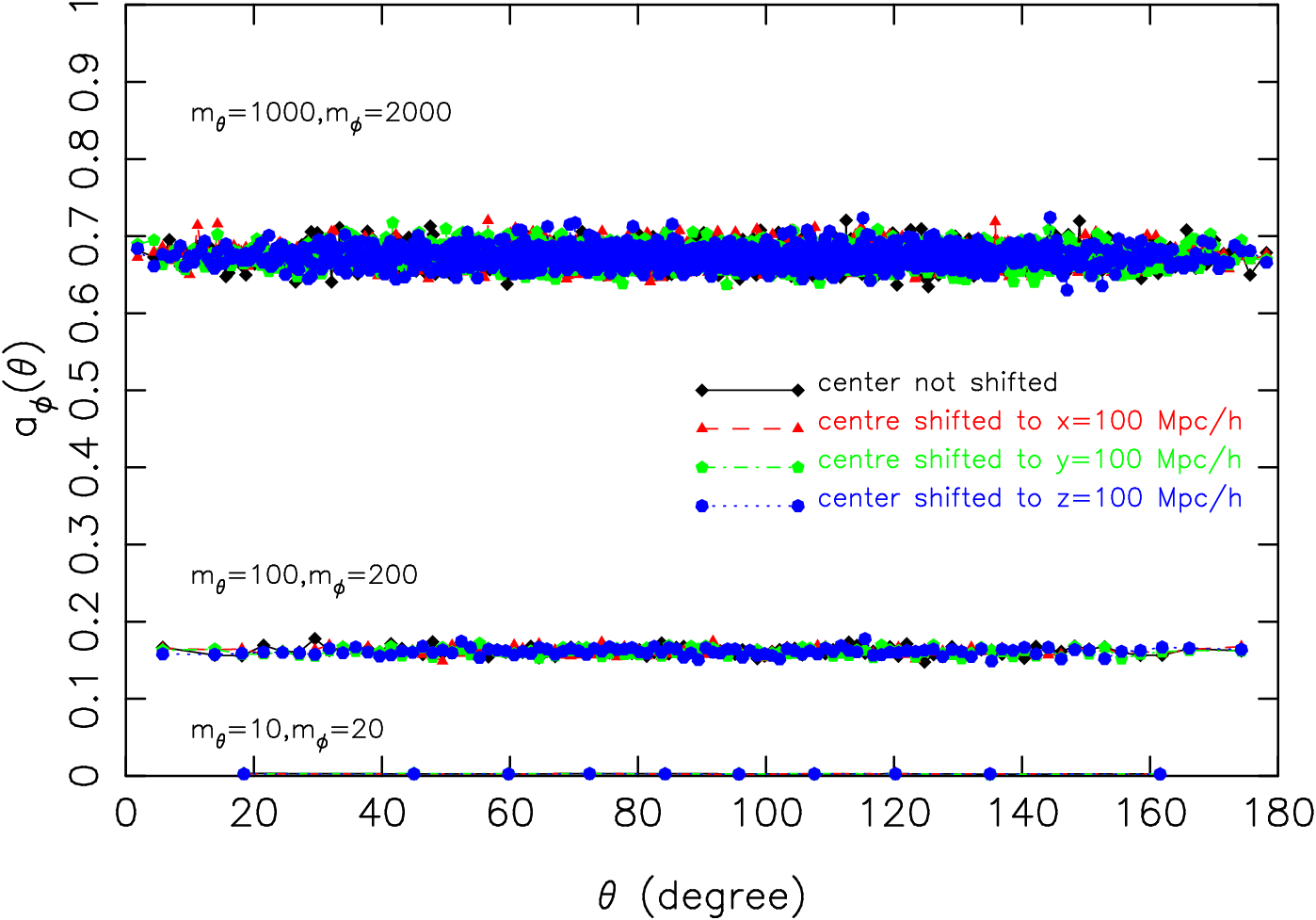}}}\\
\resizebox{9cm}{!}{\rotatebox{-90}{\includegraphics{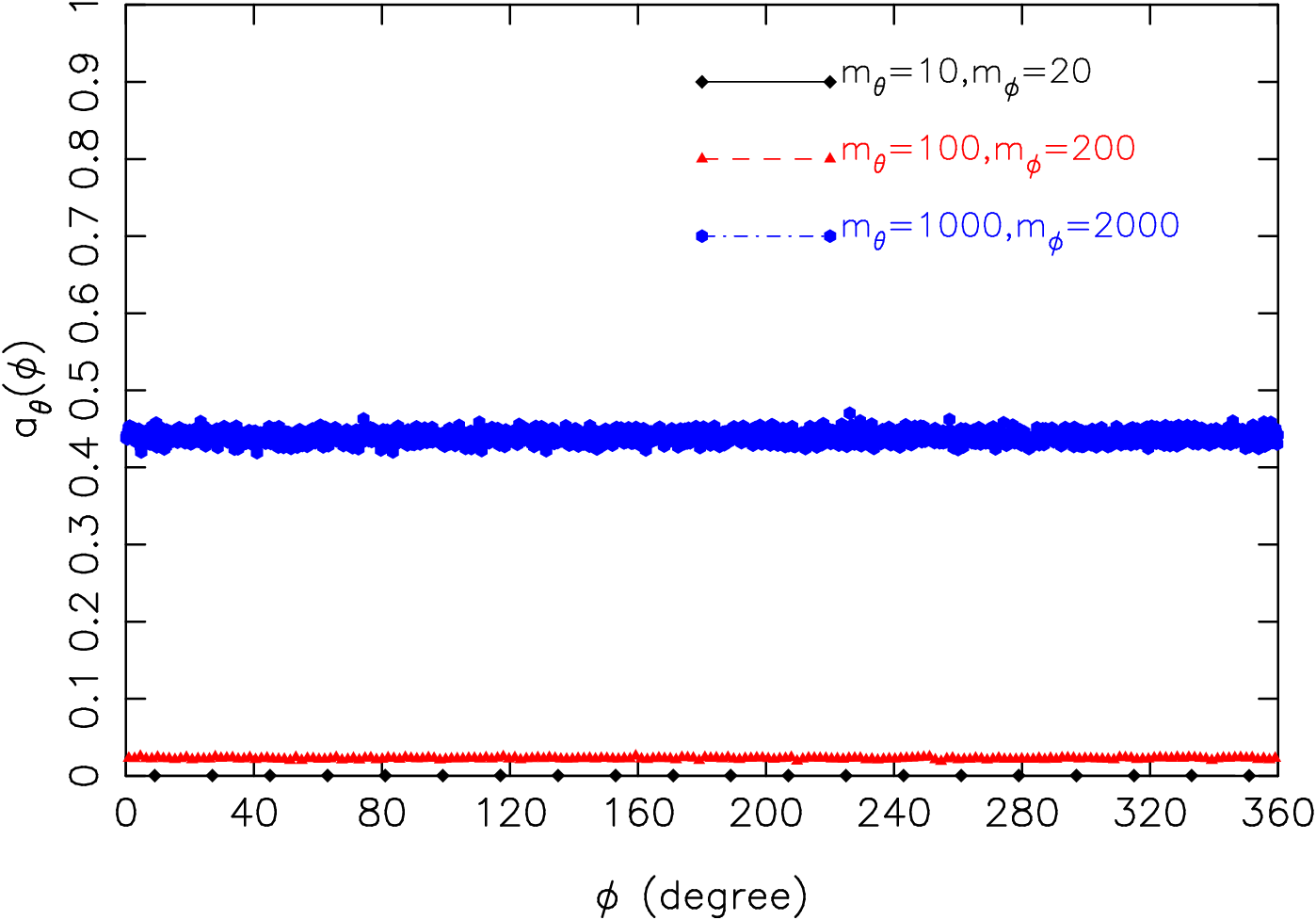}}}%
\resizebox{9cm}{!}{\rotatebox{-90}{\includegraphics{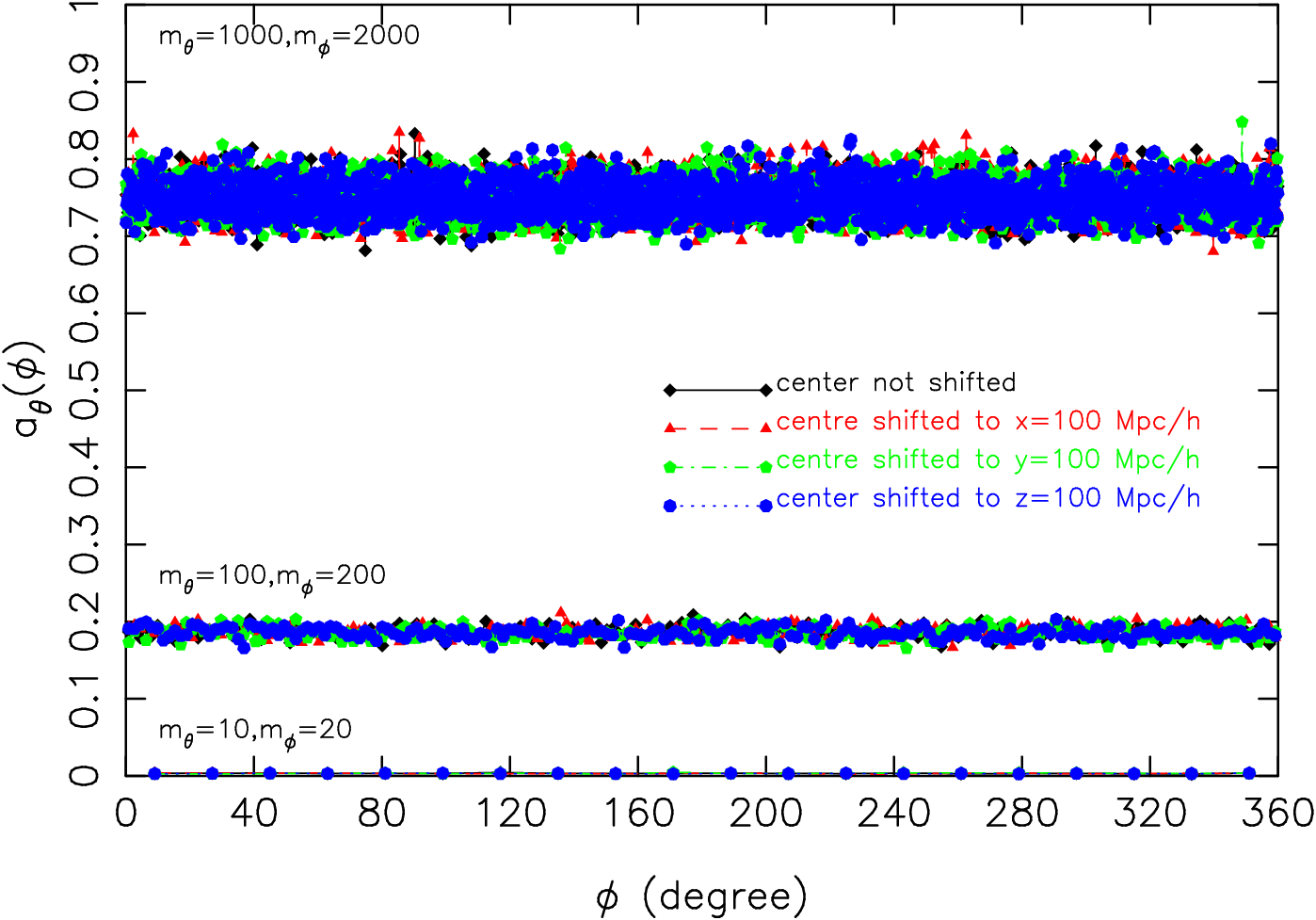}}}\\
\caption{The top left, middle left and bottom left panels show the
  measured anisotropies in a homogeneous and isotropic Poisson
  distribution as functions of $r$, $\theta$ and $\phi$ respectively
  for different choices of $m_{\theta}$ and $m_{\phi}$ as labeled in
  each panel. The top right, middle right and bottom right panels show
  the same quantities when the origin is shifted along x or y or z
  directions from the centre by $100 h^{-1} \, {\rm Mpc}$. We use
  $r_{max}=200 h^{-1} \, {\rm Mpc}$ and $r_{max}=100 h^{-1} \, {\rm
    Mpc}$ for all the panels on left and right respectively. The level
  of anisotropy seen in each panel correponds to the errors arising
  due to the discrete nature of the sampling. One needs to take into
  account these errors while testing for isotropy of any distribution
  with the same sampling rate and number of bins. The rest of our
  analysis are done at the same sampling rate and we use
  $m_{\theta}=10$ and $m_{\phi}=20$. We refer to these errors in the
  next figures.}
  \label{fig:homgp}
\end{figure*}

\begin{figure*}
\resizebox{9cm}{!}{\rotatebox{-90}{\includegraphics{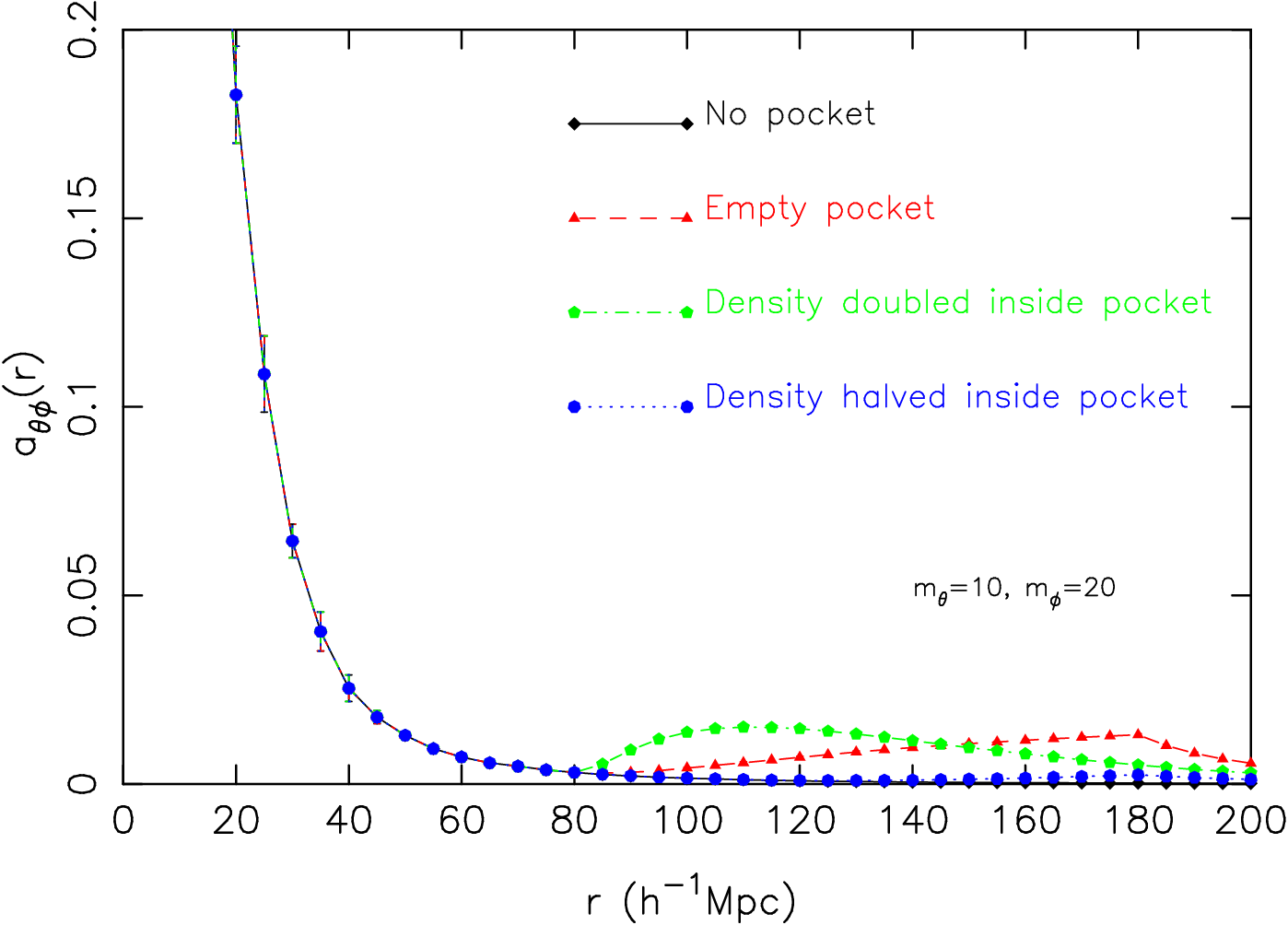}}}%
\resizebox{9cm}{!}{\rotatebox{-90}{\includegraphics{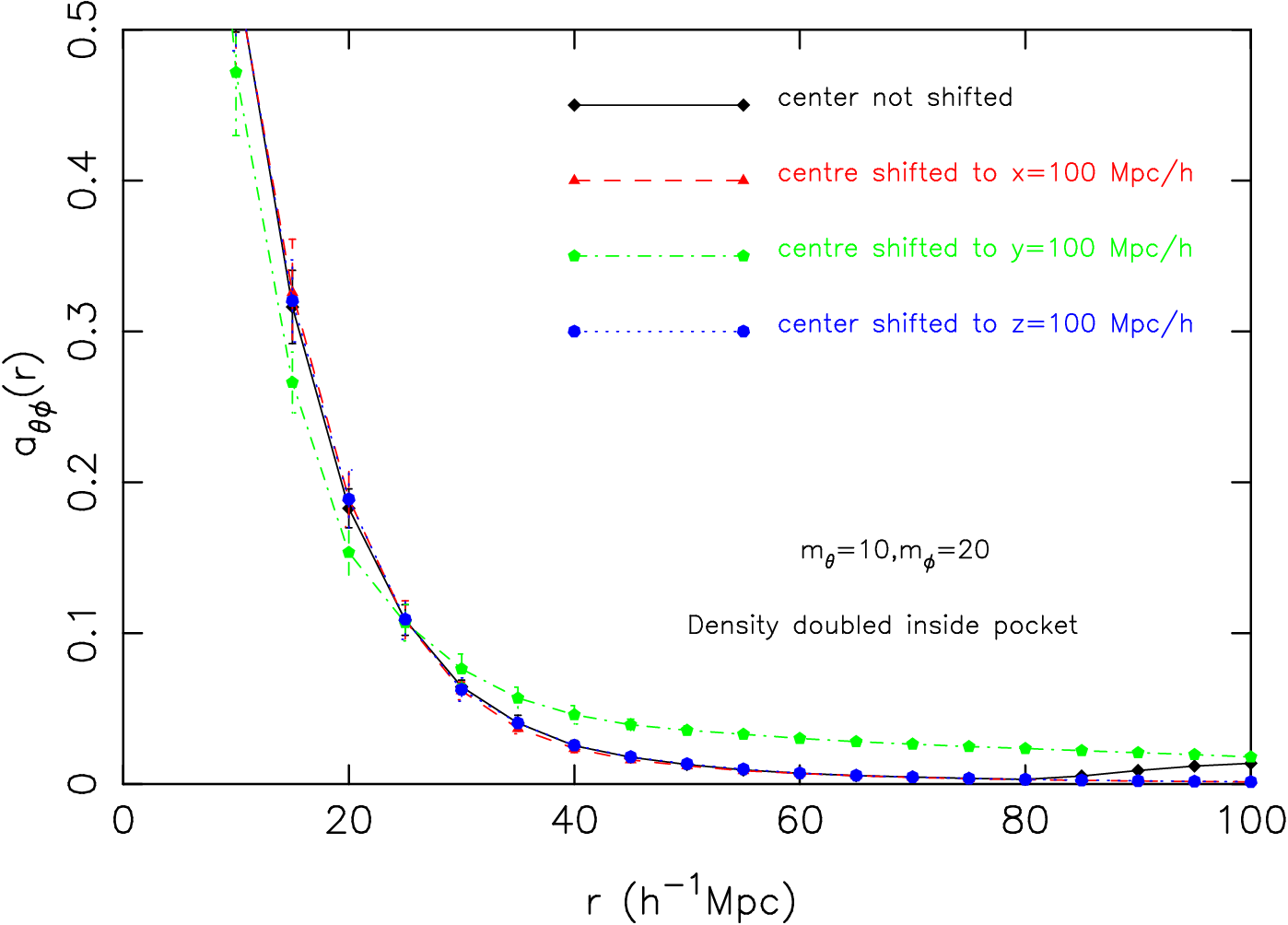}}}\\
\resizebox{9cm}{!}{\rotatebox{-90}{\includegraphics{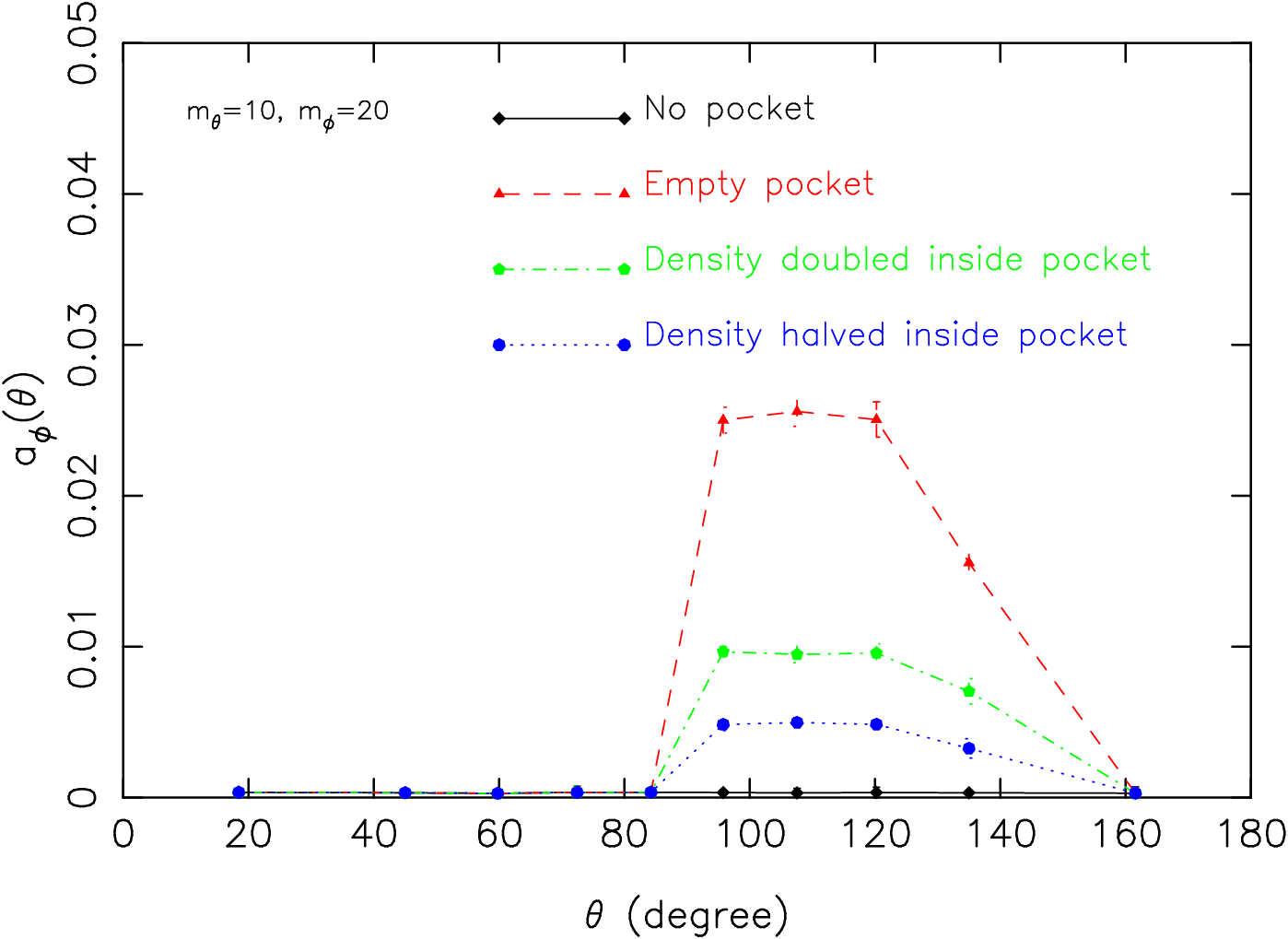}}}%
\resizebox{9cm}{!}{\rotatebox{-90}{\includegraphics{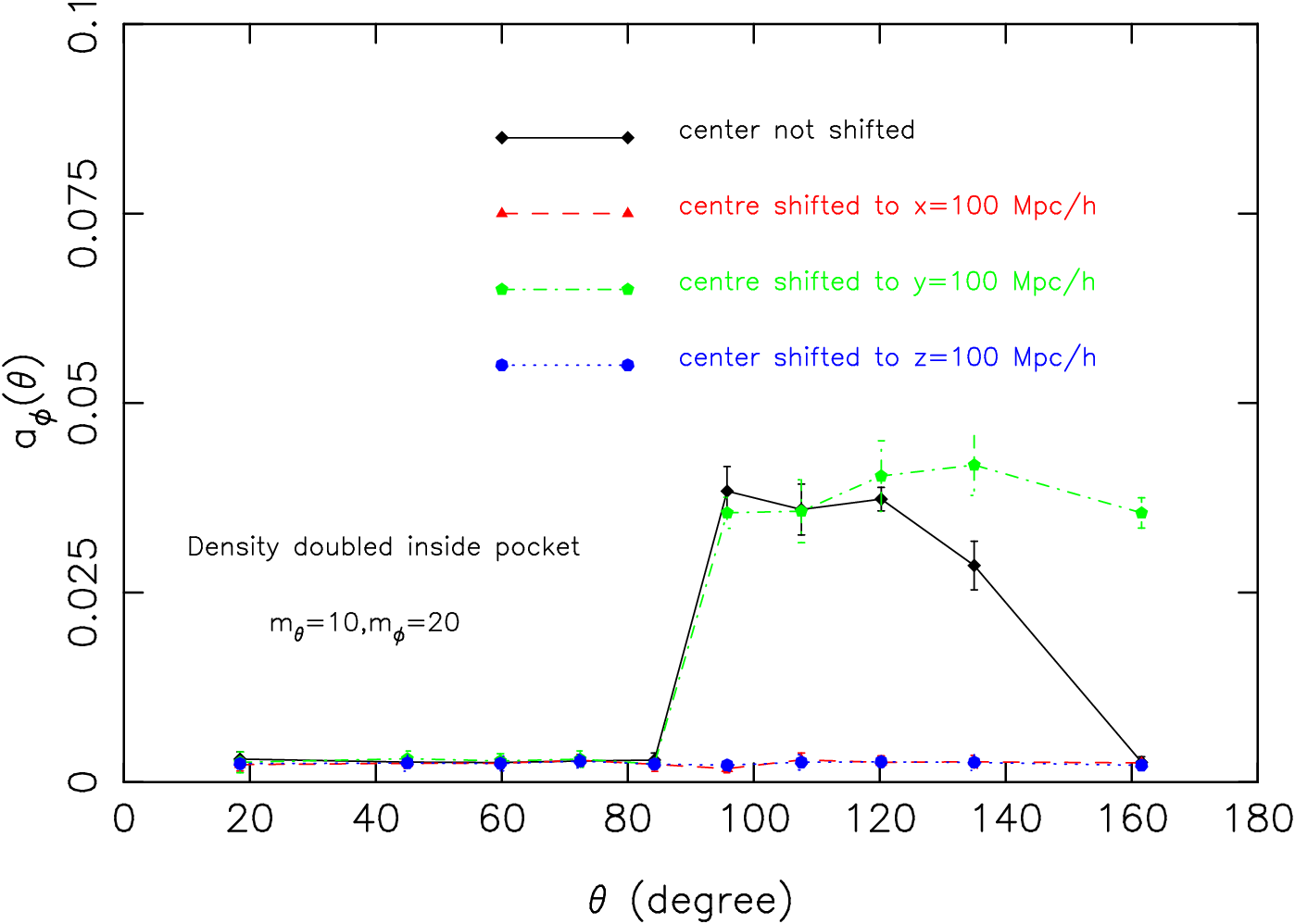}}}\\
\resizebox{9cm}{!}{\rotatebox{-90}{\includegraphics{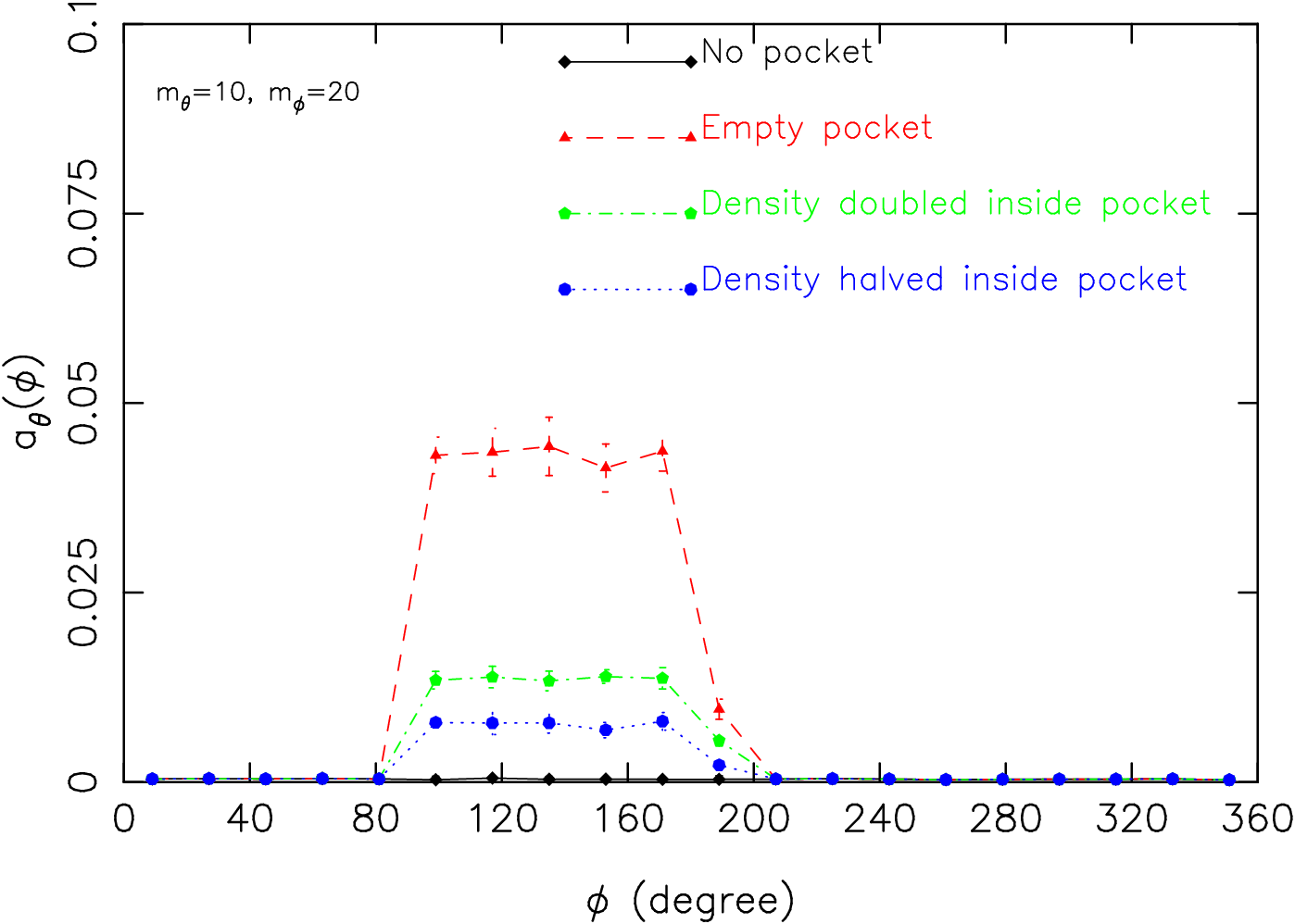}}}%
\resizebox{9cm}{!}{\rotatebox{-90}{\includegraphics{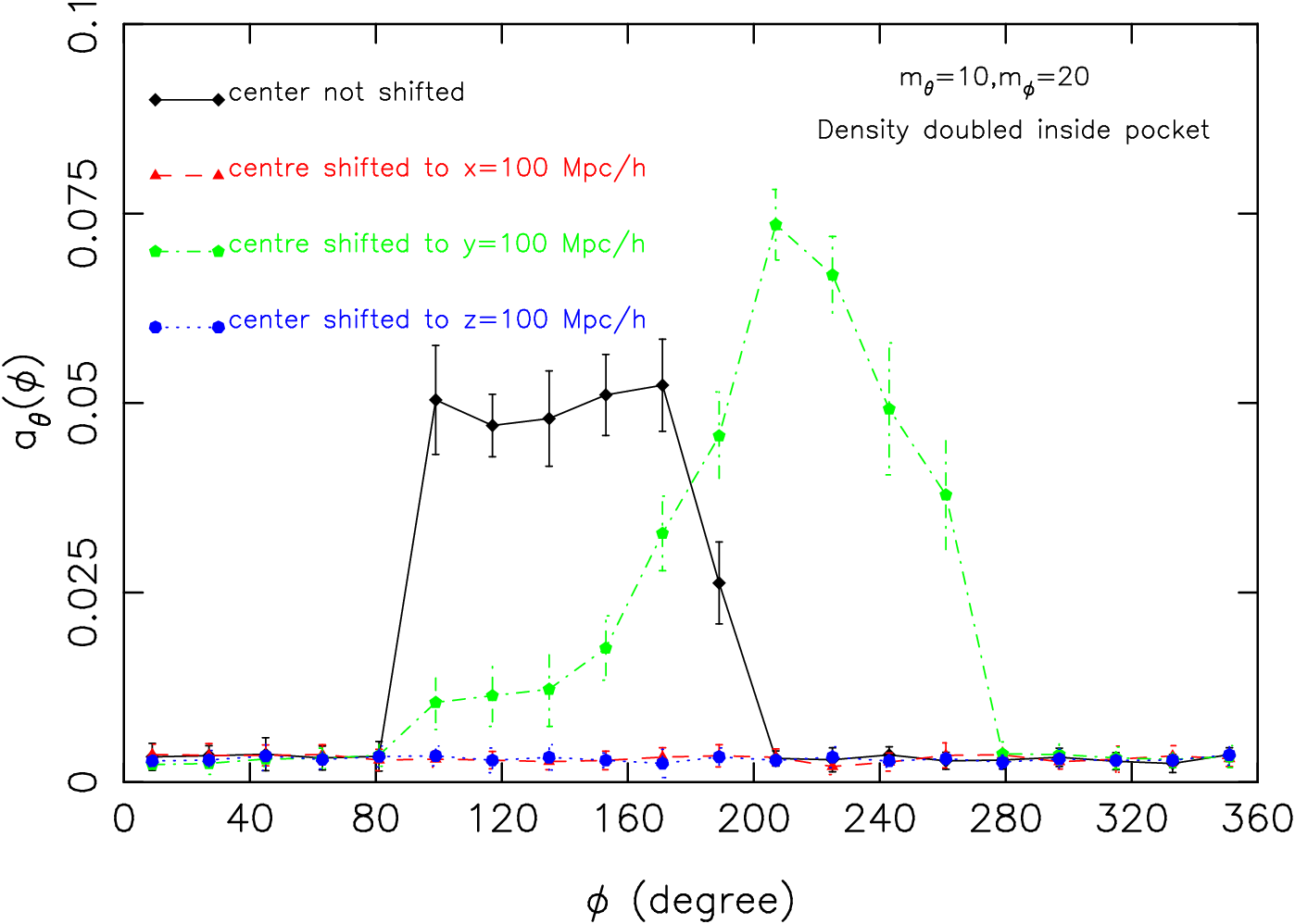}}}\\
\caption{Same as Figure 1. but for anisotropic Poisson distribution
  obtained by introducing pocket of different density in a homogeneous
  and isotropic Poisson distribution. Only the results for
  $m_{\theta}=10$ and $m_{\phi}=20$ are shown in each panel. The
  error-bars shown here in all the panels are the $1-\sigma$
  variations from the $10$ Monte Carlo realizations used in each case.
  The results for `No pocket' corresponds to the anisotropy level
  resulting from discreteness in a homogeneous and isotropic Poisson
  distribution. In all these cases the geometry of the pocket
  introduced are following: $80 h^{-1} \, {\rm Mpc} \leq r \leq 180
  h^{-1} \, {\rm Mpc}$, $90^{\circ}\leq \theta \leq 140^{\circ}$,
  $90^{\circ}\leq \phi \leq 190^{\circ}$.}
  \label{fig:pocketp}
\end{figure*}

\begin{figure*}
\resizebox{9cm}{!}{\rotatebox{-90}{\includegraphics{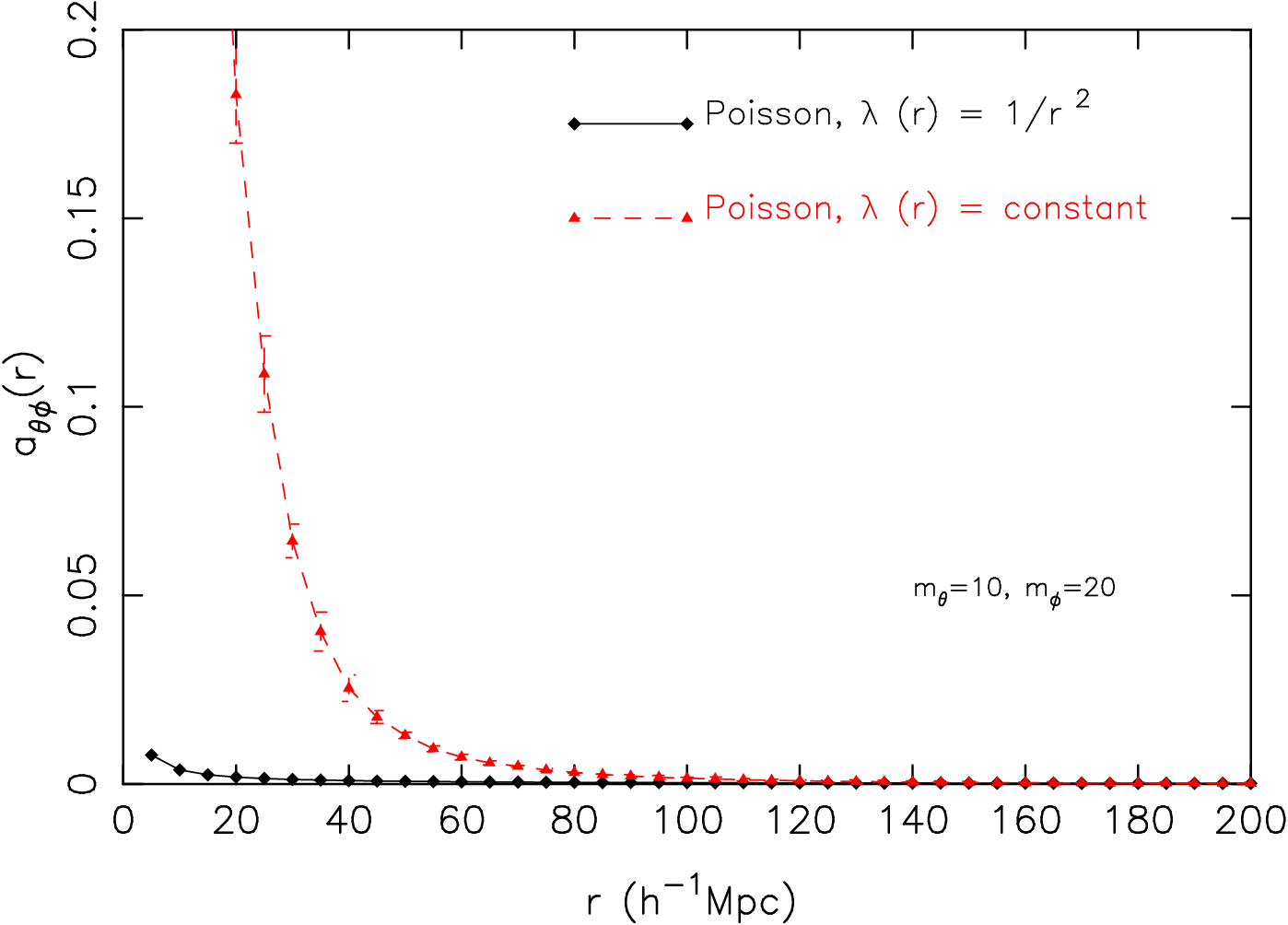}}}%
\resizebox{9cm}{!}{\rotatebox{-90}{\includegraphics{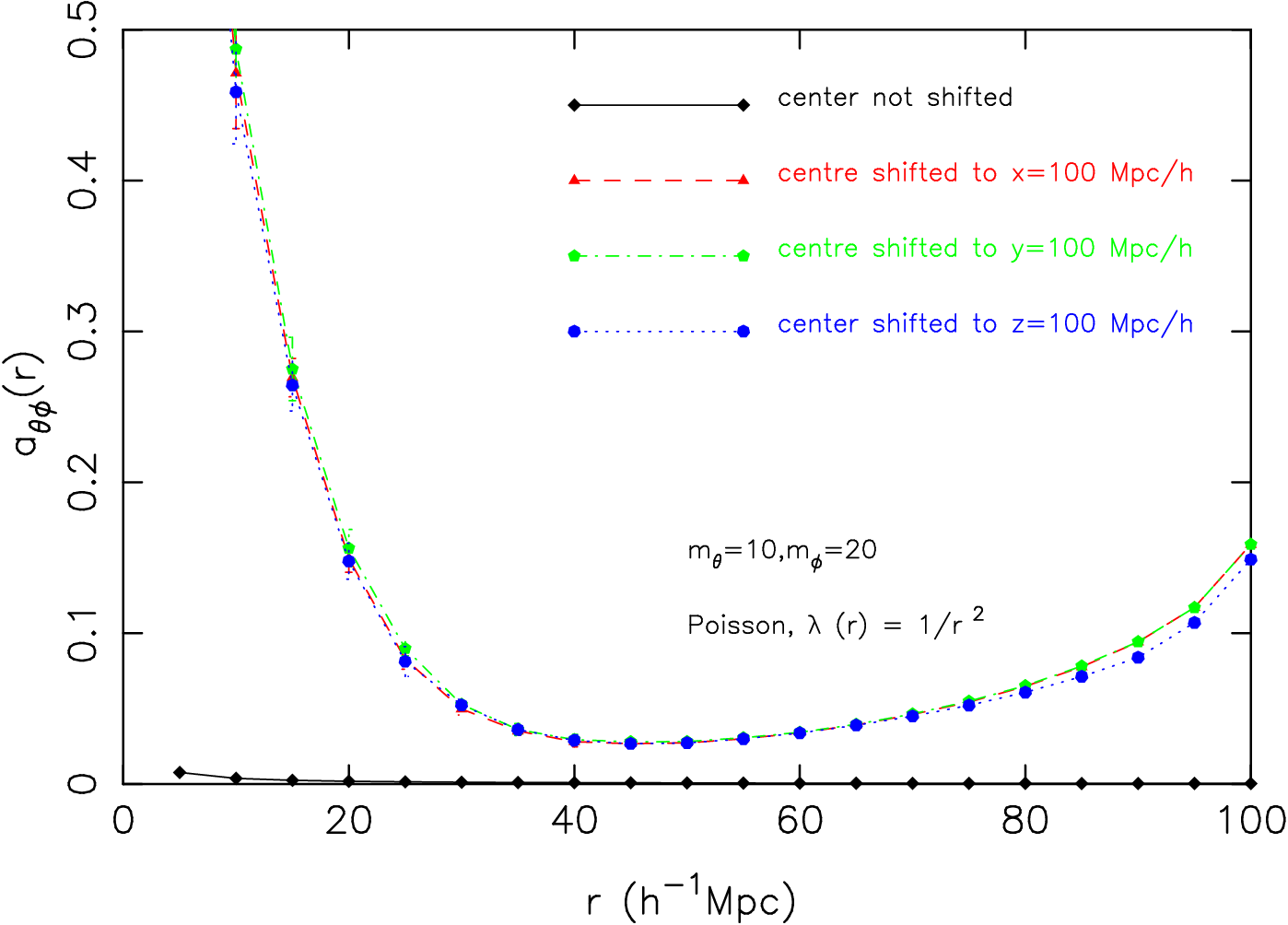}}}\\
\resizebox{9cm}{!}{\rotatebox{-90}{\includegraphics{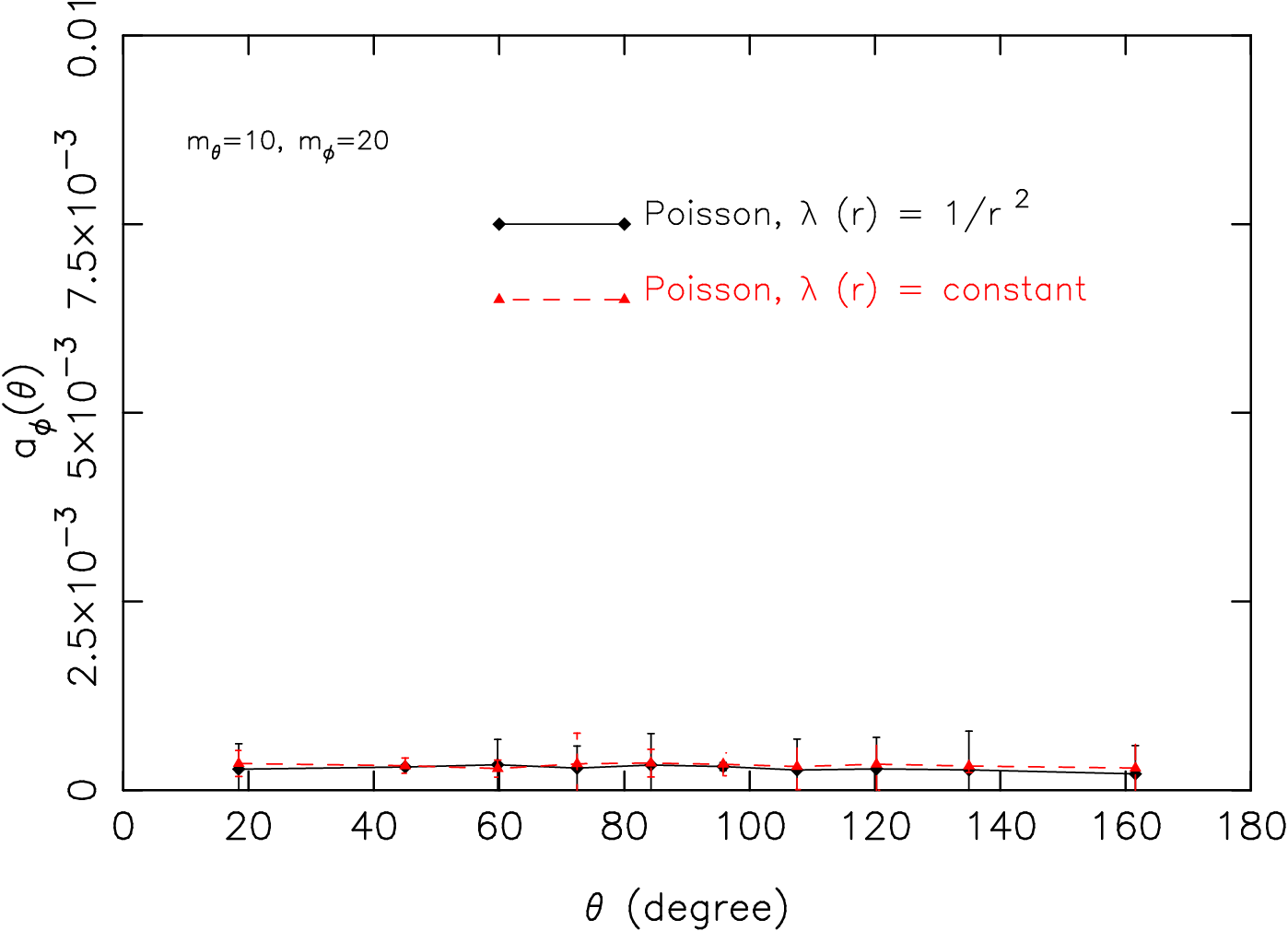}}}%
\resizebox{9cm}{!}{\rotatebox{-90}{\includegraphics{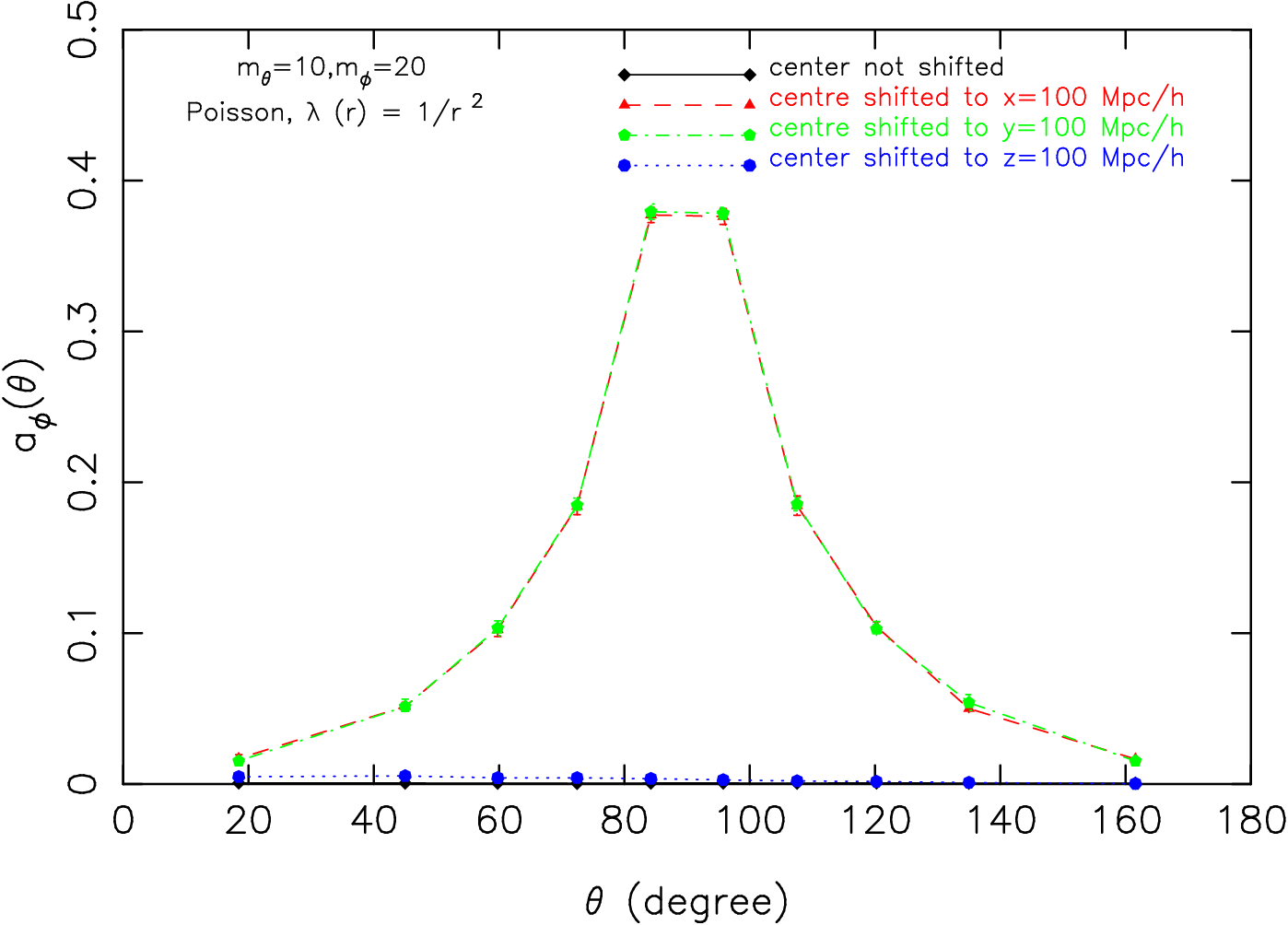}}}\\
\resizebox{9cm}{!}{\rotatebox{-90}{\includegraphics{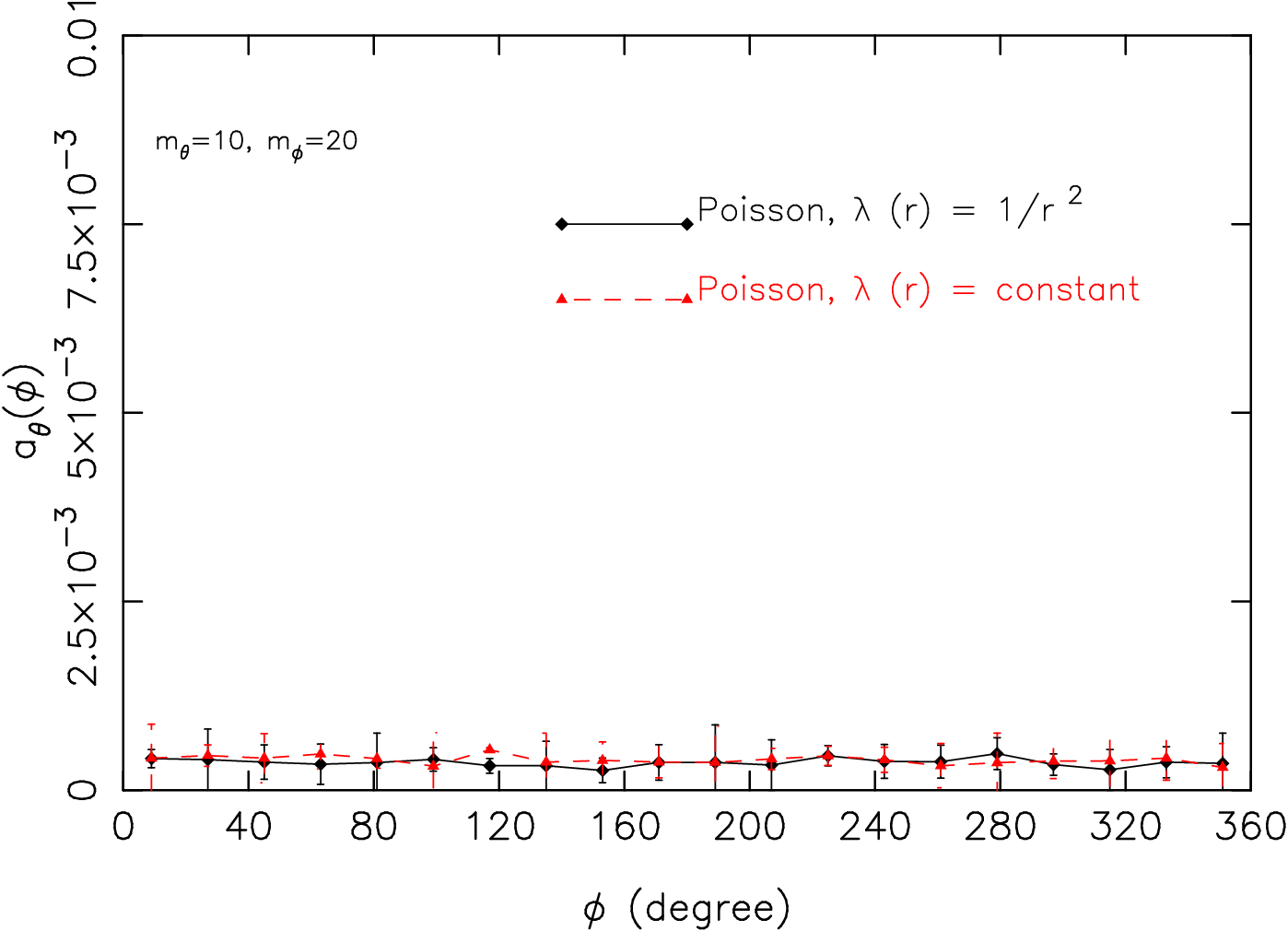}}}%
\resizebox{9cm}{!}{\rotatebox{-90}{\includegraphics{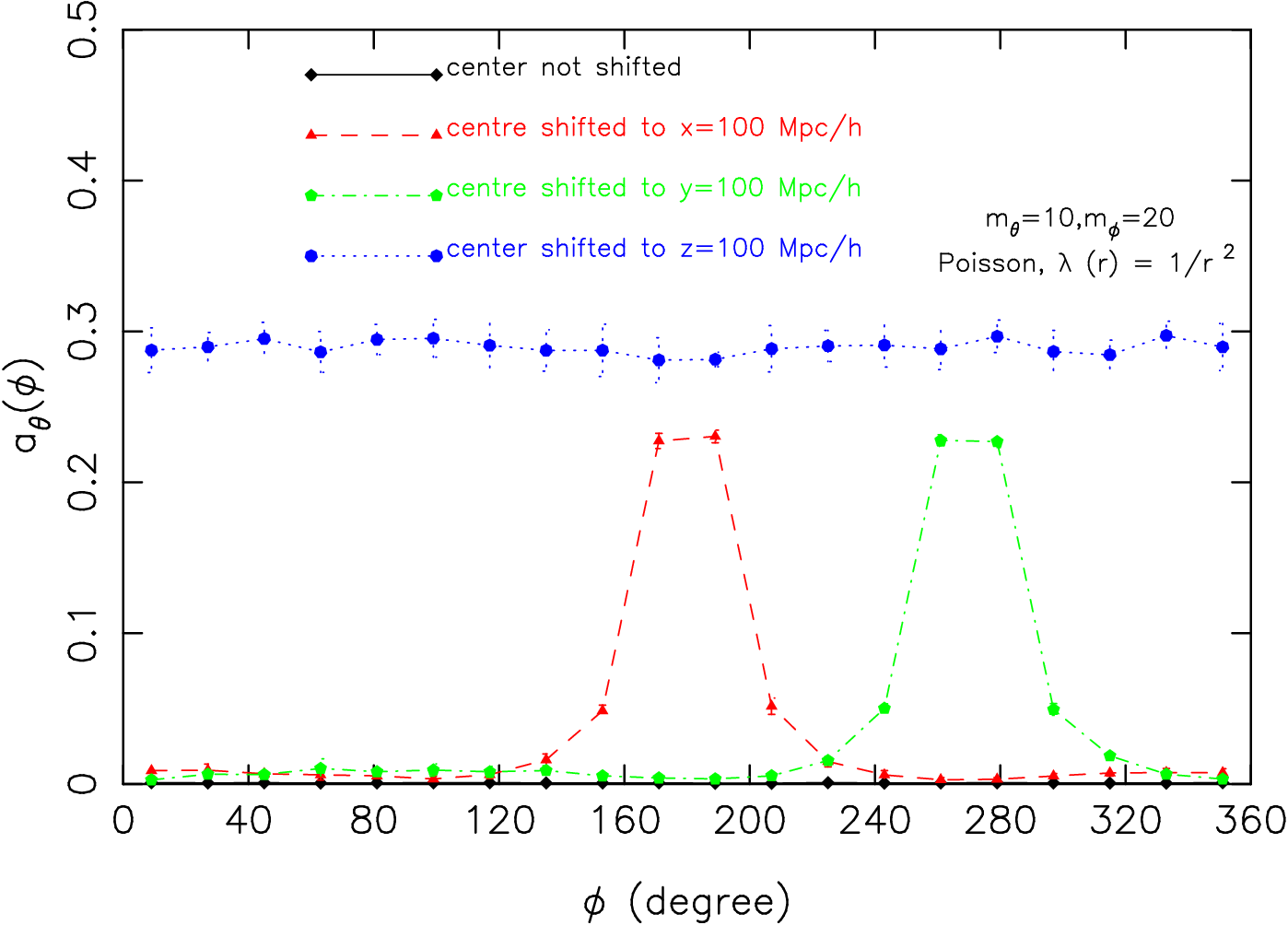}}}\\
\caption{Same as Figure 1. but for radially inhomogeneous Poisson
  distribution where density varies as $\frac{1}{r^{2}}$ from the
  centre. Only the results for $m_{\theta}=10$ and $m_{\phi}=20$ are
  shown in each panel. The error-bars shown here in all the panels are
  the $1-\sigma$ variations from the $10$ Monte Carlo realizations
  used in each case. The results for `Poisson, $\lambda(r)=$constant'
  corresponds to the anisotropy level resulting from discreteness in a
  homogeneous and isotropic Poisson distribution.}
  \label{fig:inhomrad}
\end{figure*}

\begin{figure*}
\resizebox{9cm}{!}{\rotatebox{-90}{\includegraphics{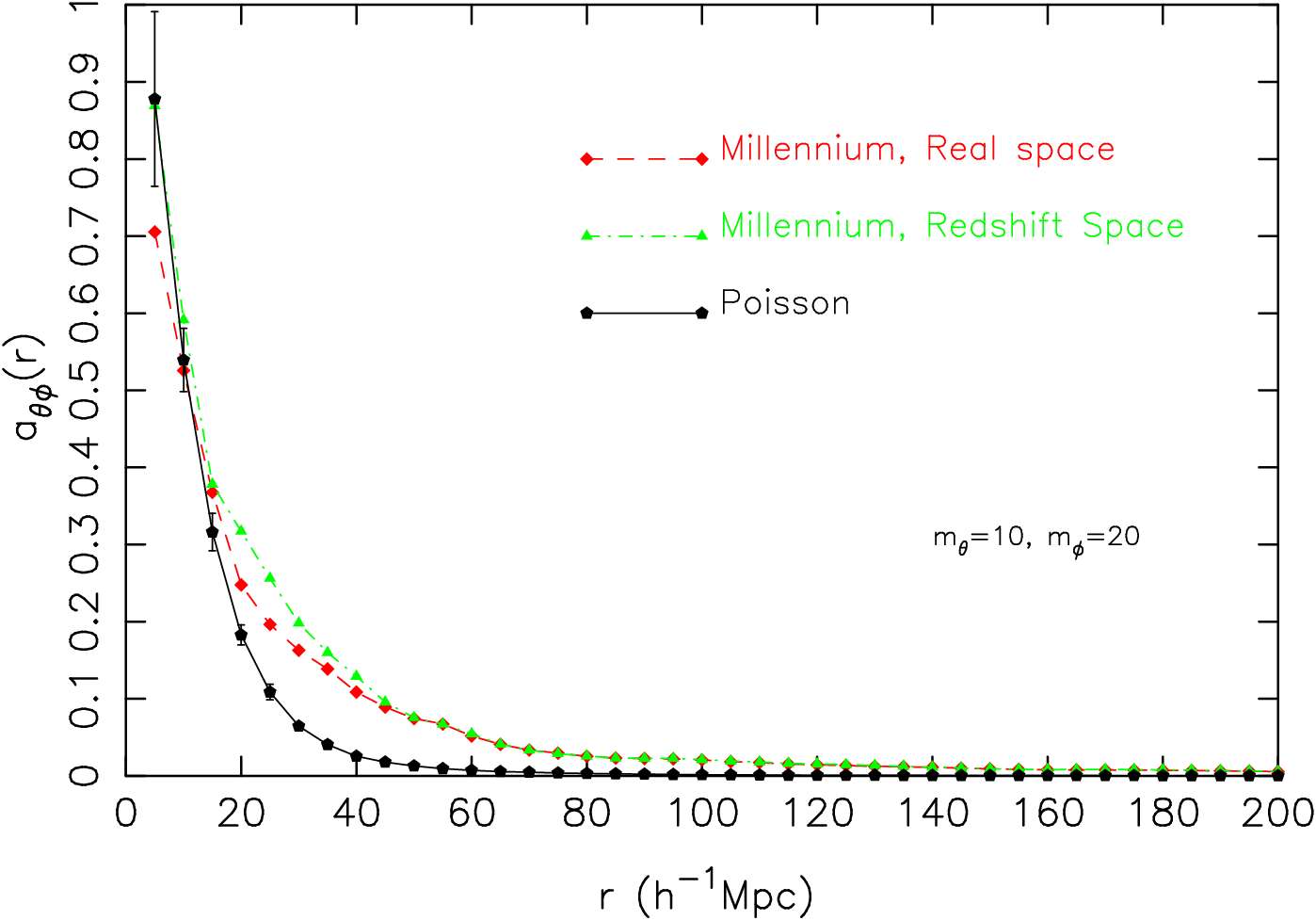}}}%
\resizebox{9cm}{!}{\rotatebox{-90}{\includegraphics{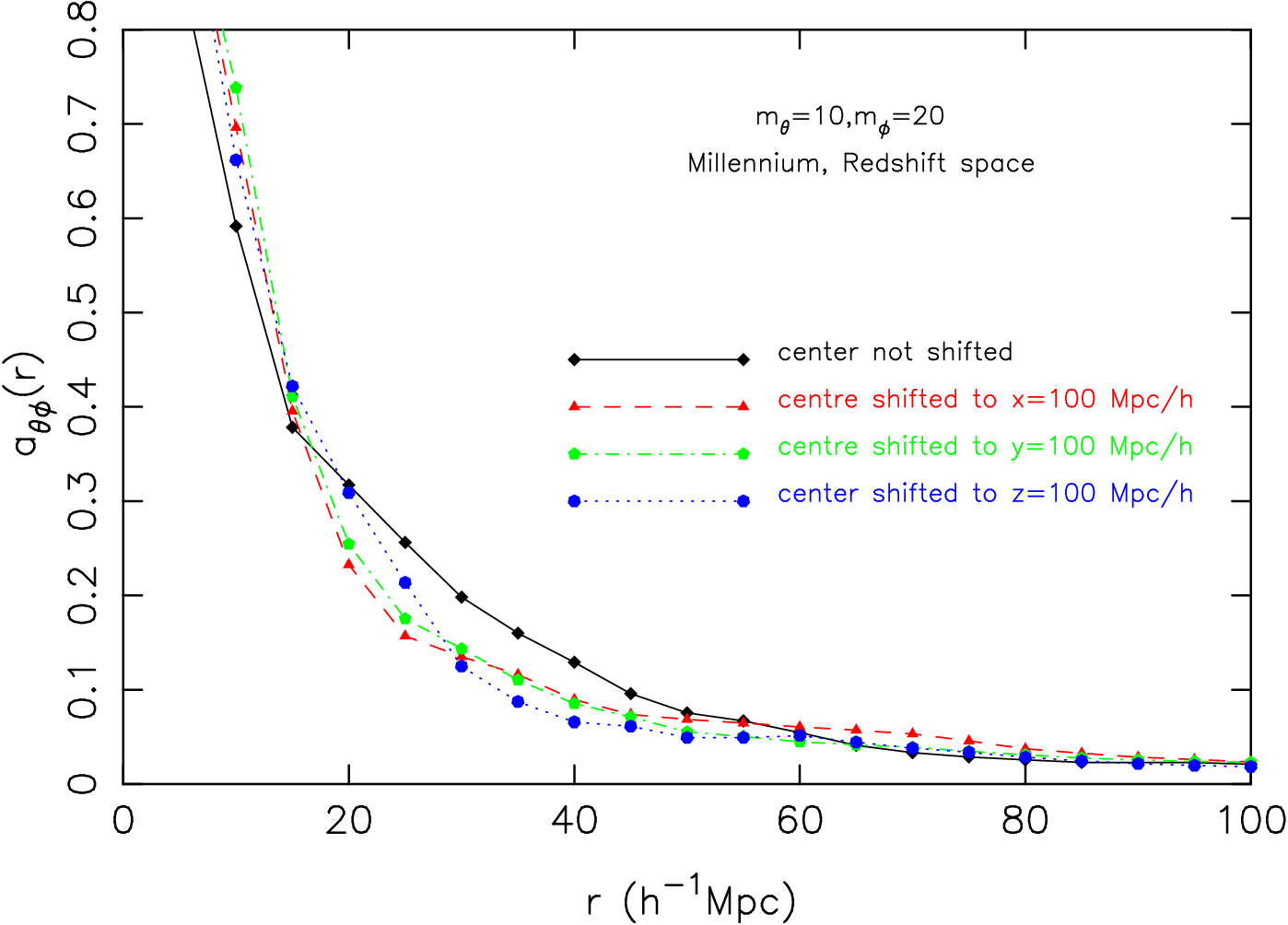}}}\\
\resizebox{9cm}{!}{\rotatebox{-90}{\includegraphics{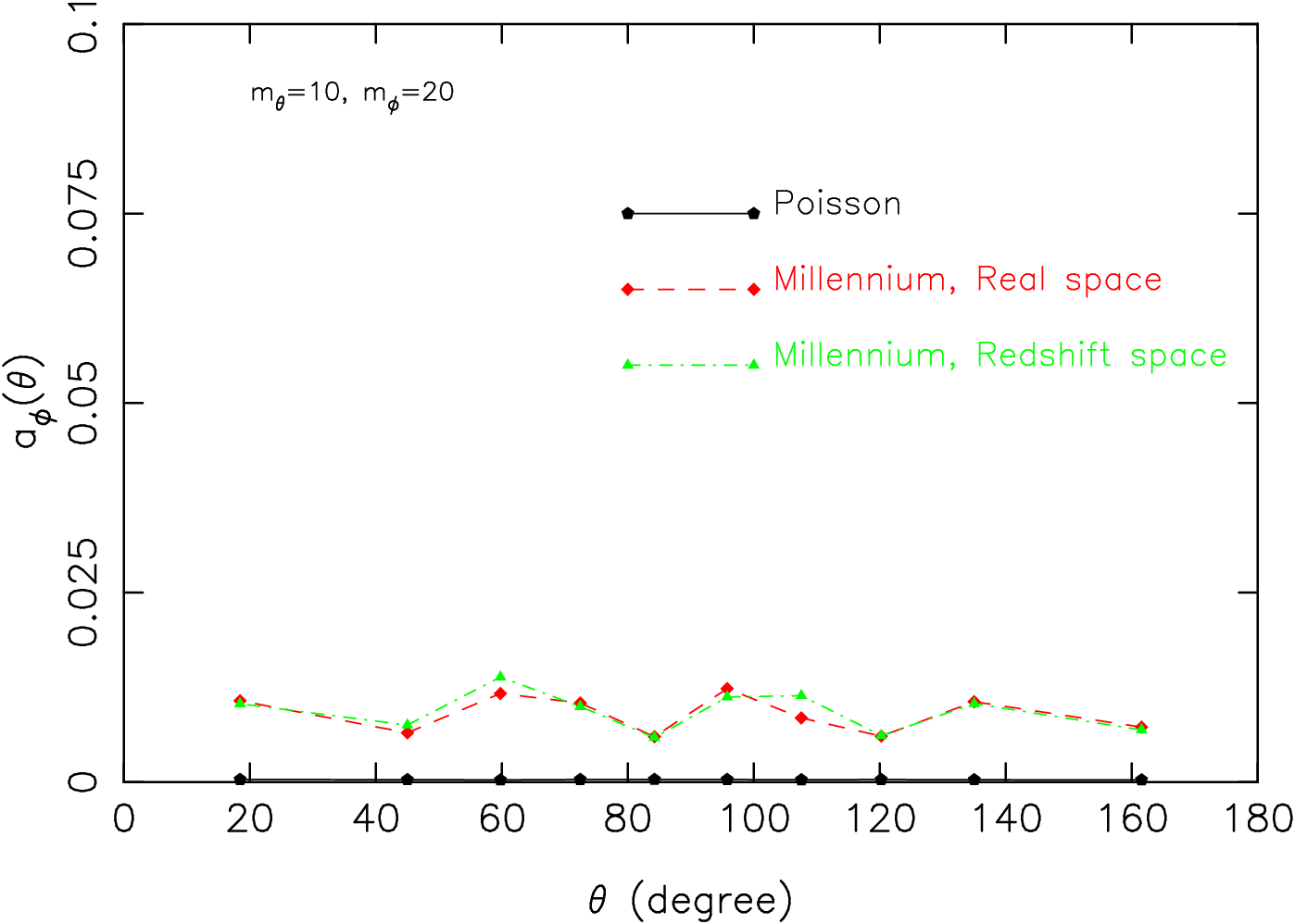}}}%
\resizebox{9cm}{!}{\rotatebox{-90}{\includegraphics{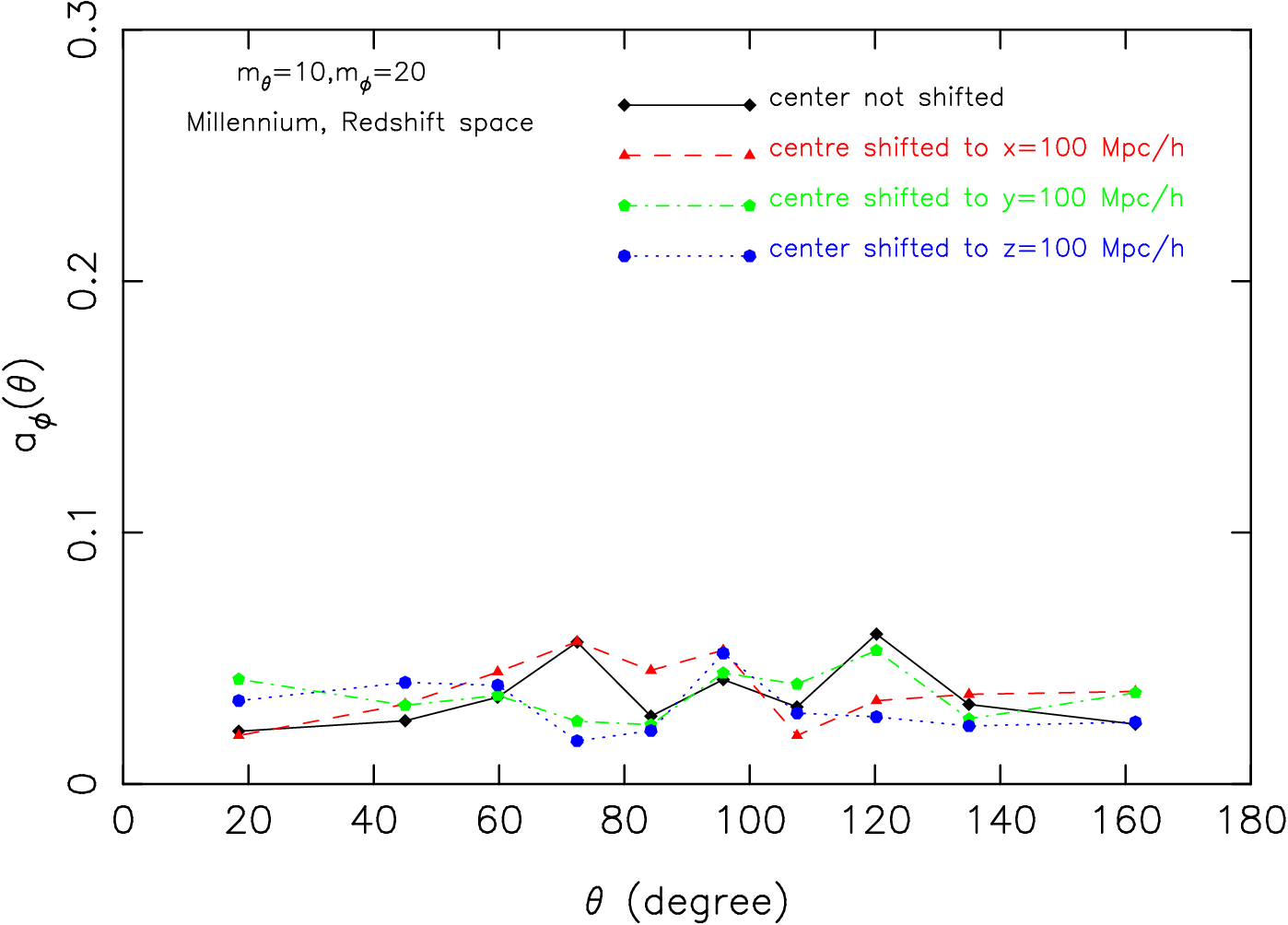}}}\\
\resizebox{9cm}{!}{\rotatebox{-90}{\includegraphics{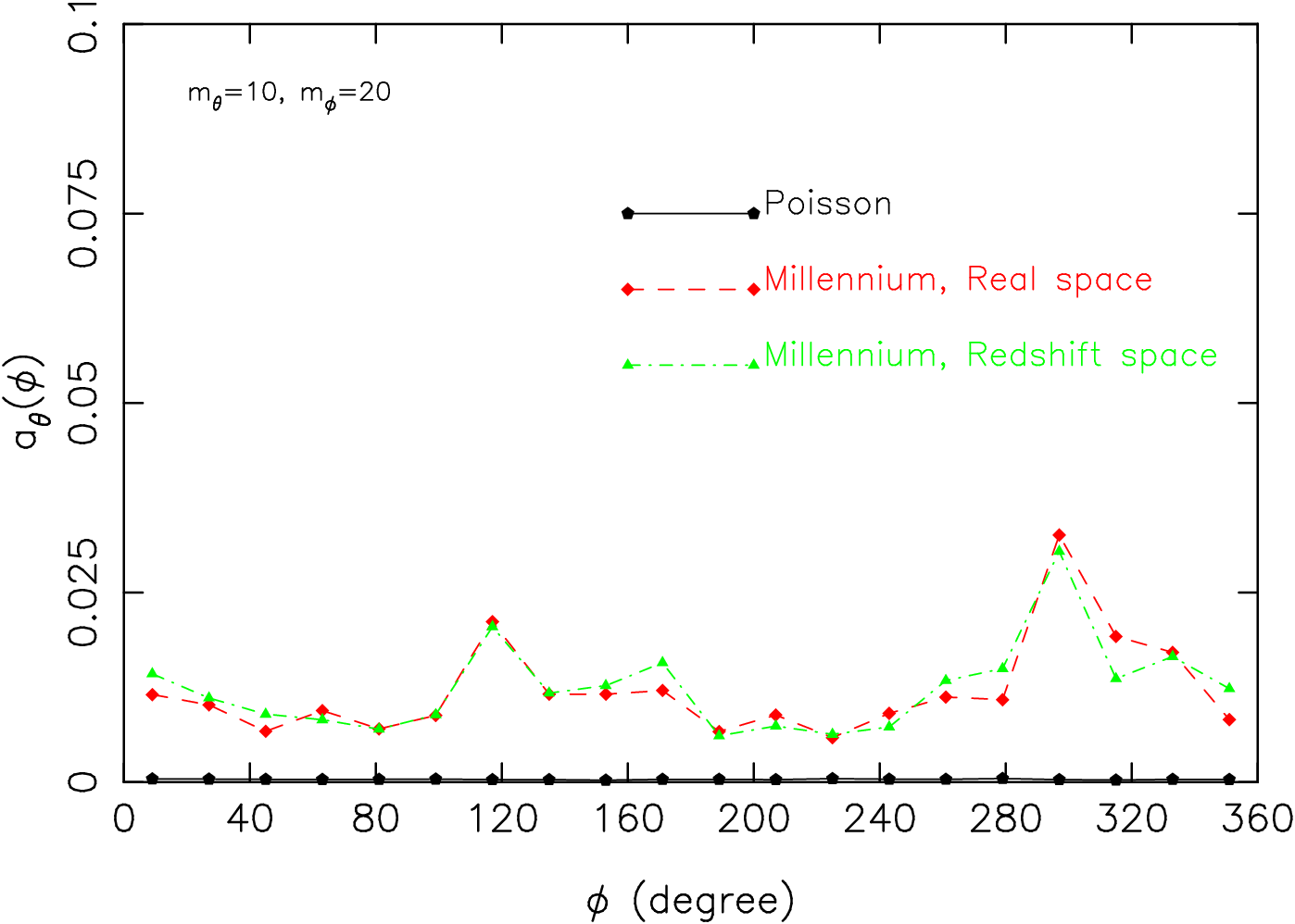}}}%
\resizebox{9cm}{!}{\rotatebox{-90}{\includegraphics{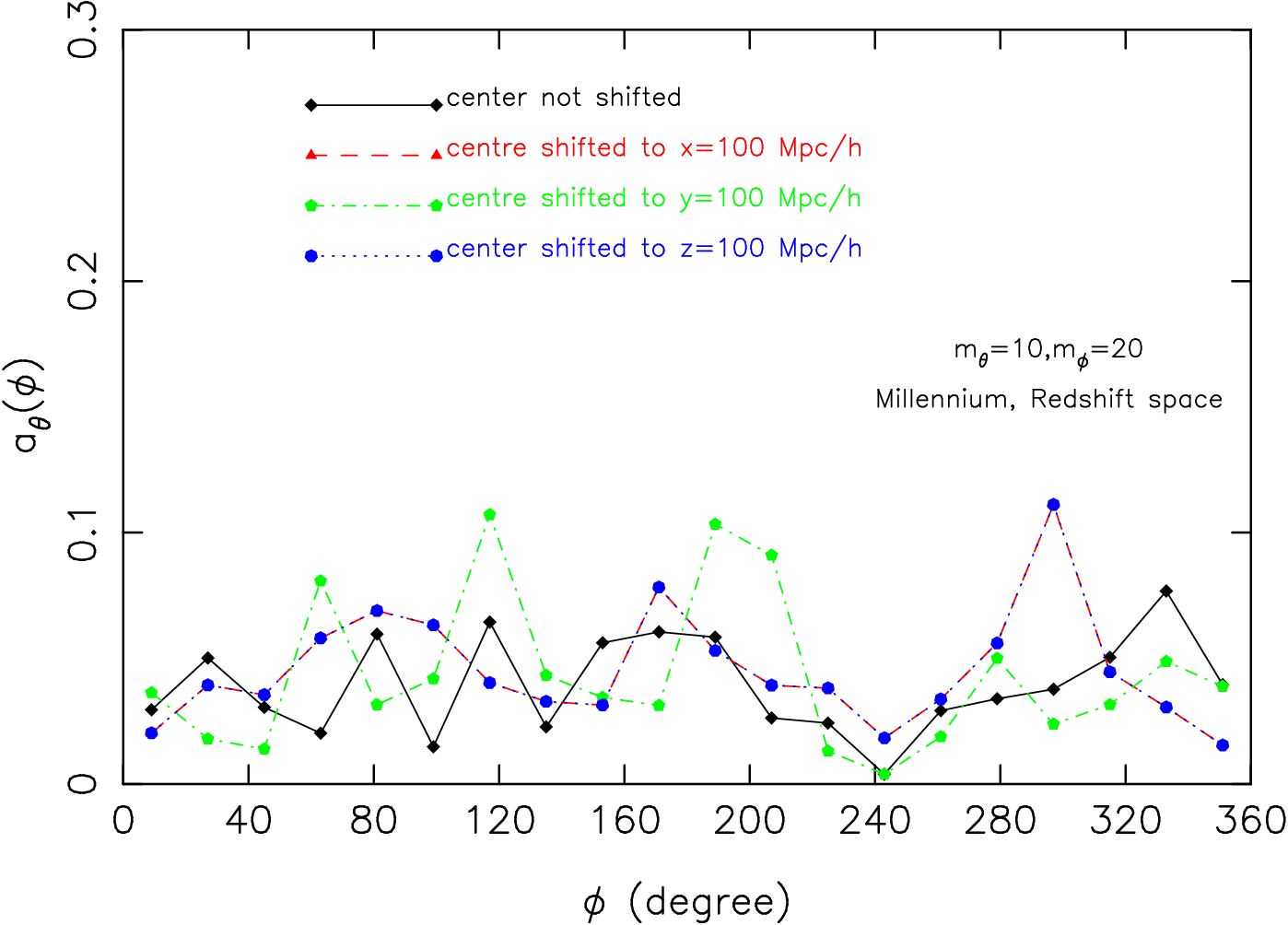}}}\\
\caption{Same as Figure 1. but for simulated galaxy distributions from
  a semi analytic galaxy catalogue from the Millennium Run
  simulation. The panels on left side compares the results for the
  real space and redshift space whereas the panels on right only show
  the results for the redshift space. The error-bars shown for the
  homogeneous and isotropic Poisson distributions are the $1-\sigma$
  variations from the $10$ Monte Carlo realizations. No error-bars are
  shown for the galaxies from the Millennium simulation as we have
  only one galaxy sample from it.}
  \label{fig:millrsd}
\end{figure*}

\section{RESULTS}
  
We show the results for the homogeneous and isotropic Poisson
distributions in \autoref{fig:homgp}. In top left, middle left and
bottom left panels we show the degree of anisotropies as a function of
$r$, $\theta$ and $\phi$ respectively. The top left panel shows the
variations of $a_{\theta\phi}(r)$ (hereafter radial anisotropy) as a
function of $r$ for different choices of $m_{\theta}$ and
$m_{\phi}$. At the smallest radius one finds a large anisotropy
arising purely from Poisson noise which gradually diminishes with
increasing radii. This result holds for all choices of $m_{\theta}$
and $m_{\phi}$ but for larger values of $m_{\theta}$ and $m_{\phi}$
the anisotropies resulting from Poisson noise are larger and persist
upto larger length scales. This arises simply because with increasing
$m_{\theta}$ and $m_{\phi}$ the solid angle bins cover smaller volumes
and hence contain fewer points within for any given $r$. The middle
left panel shows the variations of $a_{\phi}(\theta)$ (hereafter polar
anisotropy) as a function of $\theta$ for different choices of
$m_{\theta}$ and $m_{\phi}$ labeled in the panel. We note that for
$m_{\theta}=10$ and $m_{\phi}=20$ we uniformly get $a_{\phi}(\theta)
\approx 0$ for all values of $\theta$ indicating isotropy of the
distribution. But as we increase the total number of bins by $10^{2}$
times ($m_{\theta}=100,m_{\phi}=200$) and $10^{4}$ times
($m_{\theta}=1000,m_{\phi}=2000$) the anisotropies resulting from the
Poisson noise become evident. The degree of anisotropy increases with
increasing number of bins for all $\theta$ but does not change with
$\theta$ indicating a systematic behaviour as expected from Poisson
noise. We notice exactly same behaviour for $a_{\theta}(\phi)$
(hereafter azimuthal anisotropy) as a function of $\phi$ in the bottom
left panel confirming isotropy of the distribution. The entropy of a
3D Poisson point process would depend on $\lambda$ which is the
average number of points expected in a volume element in each binning
schemes. We see a variation in the radial anisotropy $a_{\phi
  \theta}(r)$ with increasing $r$ as $\lambda$ changes with increasing
radii whereas $a_{\phi}(\theta)$ and $a_{\theta}(\phi)$ do not change
with $\theta$ and $\phi$ since $\lambda$ remains constant in any
specific binning schemes. The standard deviations in the values of
$a_{\phi \theta}(r)$, $a_{\phi}(\theta)$ and $a_{\theta}(\phi)$
increase with increasing number of bins following the characteristics
of Poisson noise.

In the top, middle and bottom right panels we show how the
anisotropies vary with $r$, $\theta$ and $\phi$ respectively for
different choices of $m_{\theta}$ and $m_{\phi}$ when the origin is
shifted by $100 h^{-1} \, {\rm Mpc}$ along the x or y or z axes
without any rotation. It may be noted here that we can only probe upto
a length scale of $100 h^{-1} \, {\rm Mpc}$ under these
circumstances. We set $r_{max}=100 h^{-1} \, {\rm Mpc}$ in each of
these cases. In the top right panel we see that the
$a_{\theta\phi}(r)$ remain unaltered
when the origin is shifted by $100 h^{-1} \, {\rm Mpc}$ along the x, y
and z axes or not shifted at all indicating the isotropy of the
distribution. This result holds for each set of $m_{\theta}$ and
$m_{\phi}$. In the right middle panel we show
$a_{\phi}(\theta)$ as a function of $\theta$
when the origin is not shifted and when the origin is shifted along
three different directions. Interestingly the results overlap with
each other again pointing towards isotropy of the distribution.  One
may note here that the degree of anisotropy and the size of the
errorbars increase at a fixed choice of $m_{\theta}$ and $m_{\phi}$
due to the decrease in $r_{max}$ resulting in a smaller number of
points within the volumes analyzed. In the bottom right panel
$a_{\theta}(\phi)$ as a function of $\phi$ shows
identical behaviour showing isotropy of the distribution in $\phi$ and
rise in anisotropy with increase in the number of bins. Here the error
bars are derived from $10$ different Monte Carlo realizations in each
case. We are interested in finding out the genuine signals of
anisotropy. So keeping in mind the role of Poisson noise in increasing
the anisotropy and the size of error-bars we decide to use
$m_{\theta}=10$ and $m_{\phi}=20$ for the rest of our analysis. One
can of course safely increase $m_{\theta}$ and $m_{\phi}$ by
increasing the density of the distribution at the same time.

We introduce an empty pocket in the homogeneous and isotropic Poisson
distribution by removing all the points from the region $80 h^{-1} \,
{\rm Mpc} \leq r \leq 180 h^{-1} \, {\rm Mpc}$, $90^{\circ}\leq \theta
\leq 140^{\circ}$ and $90^{\circ}\leq \phi \leq 190^{\circ}$. This
introduces a directional asymmetry in the resulting
distribution. Subsequently we fill up the empty pocket by generating
another homogeneous and isotropic Poisson distribution within it which
has a different density than the original one. We considered two
distinct case, one in which we doubled the density and another in
which we halved the density with respect to the original
distribution. We analyze $10$ such Monte Carlo realizations in each
case.  The results are shown in \autoref{fig:pocketp}. We show the
radial anisotropy as a function of $r$ for $m_{\theta}=10$ and
$m_{\phi}=20$ in the top left panel of \autoref{fig:pocketp}. As seen
earlier in \autoref{fig:homgp} once again we see a higher degree of
anisotropy on small scales due to the Poisson noise. The anisotropy
decreases with increasing length scales when there is no pocket in the
distribution.  But in the presence of the pocket described above the
anisotropies reappear again at $80 h^{-1} \, {\rm Mpc}$ and persists
thereafter. It may be noted here that the pocket introduced radially
extends from $80 h^{-1} \, {\rm Mpc} $ to $180 h^{-1} \, {\rm Mpc} $
from the centre. We note that the signals of anisotropy in this range
of radii are less pronounced when the density of points are halved
than when doubled or when the pocket is left empty. The left middle
and left bottom panel show the polar anisotropy as a function of
$\theta$ and azimuthal anisotropy as a function of $\phi$
respectively. In the left middle panel we see a clear and distinct
bump in the polar anisotropy in the $\theta$ range $90^{\circ}$ to
$160^{\circ}$ when the pocket is introduced in the
distribution. Noticeably the bump is absent when no such pockets are
introduced. For all other $\theta$ values we uniformly get
$a_{\phi}(\theta) \approx 0$. The height of the bump quantifies the
degree of anisotropy and depends on the number density inside the
pocket. The bump clearly indicates violation of isotropy in the range
of $\theta$ values over which it appears. Interestingly the pocket
generated spans exactly the same range of $\theta$ over which the bump
appears. Consequently one can infer the size of the pocket from the
features of the bump. The fact that the the bumps extend upto $160
h^{-1} \, {\rm Mpc}$ rather $140 h^{-1} \, {\rm Mpc}$ is due to the
fact the next $\theta$ bin after $135^{\circ}$ falls only at $\sim
160^{\circ}$. The angular span of the pocket in $\theta$ can be more
accurately determined using a larger number of $\theta$ bins. The
signal of anisotropy is strongest when the pocket is left empty and
decreases when the pocket is filled with a homogeneous and isotropic
Poisson distribution with a different density than the original
one. We then compare the degree of anisotropy when the pocket is
filled with twice and half the density of the original
distribution. It is interesting to note that the signals of anisotropy
gets stronger with increasing density as it increases the disparity of
the density inside the pocket from the original one. The anisotropy
signals completely disappear when the pocket is filled with points
having same density as the original one. In the left bottom panel we
show the anisotropy as a function of azimuthal angle $\phi$ and
similarly find $a_{\theta}(\phi) \approx 0$ for all $\phi$ values
other than the range $90^{\circ}\leq \phi \leq 190^{\circ}$ where
$a_{\theta}(\phi)>0$ produce a bump in the azimuthal anisotropy. This
bump-like feature in the azimuthal anisotropy indicates violation of
isotropy in the direction $90^{\circ}\leq \phi \leq 190^{\circ}$ in
presence of the pocket. Interestingly the pocket introduced has the
same angular span in $\phi$. Clearly the bump does not appear in the
absence of any such pocket. The presence of the pocket violates the
isotropy of the distribution and is marked by the appearance of the
bumps in the radial, polar and azimuthal anisotropies. Combining these
information one can exactly infer the geometry of the pocket violating
isotropy and also infer the degree of anisotropy due to it from the
height of the bumps. Here we would like to mention that for an
arbitrary shape of the pocket it is not trivial to figure out its
exact geometry using the current method.

Next we consider only the set of distributions where the pocket is
filled with a homogeneous and isotropic Poisson distribution having
density twice than that of the original one. We again separately
measure anisotropies as a function of $r$, $\theta$ and $\phi$ from
multiple points of observation inside the distribution. We shift the
origin along x, y and z directions by $100 h^{-1} \, {\rm Mpc}$
without any rotation of the axes to have three different point of
observations. We compare our findings in the top right, middle right
and bottom right panels of \autoref{fig:pocketp}. In the top right
panel of \autoref{fig:pocketp} we show the anisotropy as a function of
$r$ for the cases where the centre is shifted along x, y or z
directions or not shifted at all. We find that the variations in
anisotropy are identical in all cases except when the origin is
shifted along y direction. This is due to the fact that the geometry
of the pocket ( $80 h^{-1} \, {\rm Mpc} \leq r \leq 180 h^{-1} \, {\rm
  Mpc}$, $90^{\circ}\leq \theta \leq 140^{\circ}$, $90^{\circ}\leq
\phi \leq 190^{\circ}$) does not affect the measurements in any of
these cases other than when the shift is applied along y
direction. This shows the existence of a preferred direction which
clearly violates isotropy. This becomes even clearer in the right
middle and right bottom panels of \autoref{fig:pocketp}. In the right
middle panel we see that a bump appears between $90^{\circ}$ to
$160^{\circ}$ when the centre is not shifted. The pocket radially
spans from $80-180 h^{-1} \, {\rm Mpc}$ and hence overlaps with the
current measurement producing an anisotropy signal in the appropriate
range of $\theta$. On the other hand the pocket is completely excluded
from the measurements when the centre is shifted along x or z
directions showing a near uniform very small signal of anisotropy for
all $\theta$ values. This small signal of anisotropy arises due to the
Poisson noise resulting simply from the reduction in the radial
extensions of the volume elements used in the measurements. But when
we shift the origin along y direction by $100 h^{-1} \, {\rm Mpc}$ the
measurements include major part of the pocket showing anisotropy in
the relevant range of $\theta$. As there are no rotations of the axes,
the bump appears exactly at $\theta=90^{\circ}$ irrespective of
whether the centre is shifted along y direction or not shifted at
all. But the shifted position of the origin redefines the geometry of
the pocket which changes the upper limit of $\theta$ for the pocket
when the centre is shifted along y direction. We see very similar
results in the right bottom panel of \autoref{fig:pocketp} where
anisotropies are shown as a function of $\phi$. We again find presence
of bumps over the appropriate range of $\phi$ values when the centre
is shifted along y directions or not shifted at all whereas the
distribution appears to be isotropic when the origin is shifted along
x or z directions. These differences clearly indicate the presence of
anisotropies in the distribution. The error-bars shown here in all the
panels are the $1-\sigma$ variations from the $10$ Monte Carlo
realizations used in each case.

For the inhomogeneous Poisson distributions the density only varies
radially as $\frac{1}{r^{2}}$ from the centre. This preserves the
isotropy of the distribution about the centre but isotropy is violated
for all other points. We want to test if our method can capture these
expected behaviours for such distributions. The results are shown in
different panels of \autoref{fig:inhomrad}. In top left panel we show
the anisotropy as a function of radial distance $r$ when the origin is
located at the centre of the spherical volume. When we compare the
anisotropy in the homogeneous and radially inhomogeneous Poisson
distributions we find that the former shows a higher degree of
anisotropy than the later. The inhomogeneous Poisson distribution
considered here has a radial variation in density as $\frac{1}{r^{2}}$
from the centre. As a result the number density of points are
significantly higher at smaller radii in the inhomogeneous Poisson
distribution as compared to the homogeneous one. This leads to
significant reduction in the Poisson noise at smaller radii where it
is considered to be more dominant. At larger radii the situation would
be just opposite but as the number counts are cumulative in our method
the distribution would remain isotropic as expected despite the radial
decrease in density. In the middle left and bottom left panels of
\autoref{fig:inhomrad} we show the anisotropies as function of
$\theta$ and $\phi$ respectively. In both cases we have used
$r_{max}=200 h^{-1} \, {\rm Mpc}$. We find that as expected the
distributions are found to be highly isotropic from the centre.

The top right panel of \autoref{fig:inhomrad} shows the anisotropies
as a function of $r$ when the origins are shifted from the centre by
$100 h^{-1} \, {\rm Mpc}$ along x or y or z direction. We see a large
anisotropy at smaller radii which decreases with increasing radii due
to the relative increase in the number counts and again increases
afterwards due to large disparity in the density at the central and
peripheral regions. The variations in anisotropies are identical when
the origin is shifted from the centre along x, y or z directions due
to the identical variations in density along all radial directions but
they are noticeably different from the results obtained without
shifting the origin from the centre. This demonstrates the anisotropic
nature of the distribution.

The middle right panel of \autoref{fig:inhomrad} show the anisotropies
as a function of $\theta$ when the origin is shifted in various
directions.  It is interesting to note that when we shift the origin
along z direction the distribution appears to be isotropic in $\theta$
as it would appear without any shift at all. The relative Shannon
entropy is measured across all the $\phi$ bins at each $\theta$
values. The polar angle $\theta$ is defined with respect to the z axis
and all the $\phi$ bins at a specific $\theta$ value are located at
the same distance from the centre. Since the density only changes in
the radial directions the distribution would appear isotropic in all
$\phi$ directions for each $\theta$ when the origin is shifted in the
z direction. The situations are not analogous when the origin is
shifted in the x or y direction. In both the cases different $\phi$
bins at any given $\theta$ value are located at different distances
from the centre leading to variations in their densities. This is true
for all $\theta$ values albeit with a different degree of variation.
The degree of variation across all the $\phi$ bins is expected to peak
at $\theta=90^{\circ}$ as it would encompass largest variation in the
radial distances among the $\phi$ bins. Interestingly our method
capture these predictable behaviours of anisotropies as a function of
$\theta$ quite well. 

Finally in the bottom right panel of \autoref{fig:inhomrad} we show
the anisotropies as a function of $\phi$. Here we estimate the
relative Shannon entropy utilizing the information across all the
$\theta$ bins at each $\phi$ values. When the origin is shifted in z
direction the azimuthal angle $\phi$ is redefined in the shifted x-y
plane which lies at a fixed distance $100 h^{-1} \, {\rm Mpc}$ from
the centre. Consequently different $\theta$ bins at any given $\phi$
value are at different distances from the centre leading to variations
in their densities but the degree of variations across the different
$\theta$ bins for each $\phi$ value would be exactly same as the
$\theta$ bins cover same variations in their radial distances. In the
bottom right panel we see a constant degree of anisotropy across all
the $\phi$ values when the origin is shifted along z direction. On the
other hand when the origin is shifted along x or y direction the
available $\theta$ bins at each $\phi$ value are located at different
distances from the centre. As a result the different $\theta$ bins at
each $\phi$ values would exhibit different number density depending on
their distances from the centre. But the degree of variations would
not be same for all the $\phi$ bins as it depends both on the distance
range covered by the corresponding $\theta$ bins as well as if those
$\theta$ bins lie towards or away from the centre. Clearly these
variations are expected to peak at $\phi=180^{\circ}$ and
$\phi=270^{\circ}$ for shift along x and y directions respectively. We
exactly recover these predictable behaviours in the bottom right
panel. Differences in the anisotropies with shifts and without shift
clearly indicate that the distribution is anisotropic in nature. These
results together indicate that our method is not only able to sense
the anisotropies but also can capture the nature of anisotropies
present in a distribution.

In \autoref{fig:millrsd} we investigate the anisotropies resulting
from the redshift space distortions using a semi analytic galaxy
catalogue from the Millennium Run simulation. In the top left panel of
this figure we compare the anisotropies as a function of radius $r$
for the simulated galaxy distributions in real and redshift
space. Both distributions show anisotropies on small scales which
partly arise due to inevitable Poisson noise. The degree of anisotropy
gradually decreases with increasing radii. Noticeably at smaller radii
the degree of anisotropy in both the distributions are higher as
compared to a homogeneous and isotropic Poisson distribution
indicating the presence of additional sources of anisotropy other than
the Poisson noise. We see that on smaller radii the redshift space
distribution of the simulated galaxies are more anisotropic than its
real space counterpart. This is most likely caused due to the
elongation of virialized clusters, compression of large scale
overdensities and elongation of large scale underdensities along the
line of sight in redshift space. The results clearly indicate that
distribution of galaxies inside the cosmic web is not isotropic even
in real space and redshift space distortions only enhance these
anisotropies further. The differences between the real space and
redshift space anisotropies cease to exist beyond $40 h^{-1} \, {\rm
  Mpc}$ and both the anisotropies become almost indistinguishable from
that observed in homogeneous and isotropic Poisson distributions at
$\sim 140 h^{-1} \, {\rm Mpc}$. It may be worth mentioning here that
in an earlier work \citep{pandey15} we find that observed galaxy
distribution in the SDSS DR12 appears to be homogeneous on scales
above $140 h^{-1} \, {\rm Mpc}$. In the middle left and bottom left
panels of \autoref{fig:millrsd} we show the anisotropies as a function
of $\theta$ and $\phi$ respectively. In both these panels we find that
the real and redshift space distributions are equally isotropic in
$\theta$ and $\phi$. A small signal of anisotropy exist for both real
and redshift space distributions of the simulated galaxies which
separates them from identical homogeneous and isotropic Poisson
distributions. The real and redshift space distributions demonstrate
equal degree of isotropy both in $\theta$ and $\phi$ due to the fact
that the compression and elongation of overdense and underdense
regions are symmetric along the line of sight. We use $r_{max}=200
h^{-1} \, {\rm Mpc}$ for all the results shown in all the panels on
left of \autoref{fig:millrsd}. 

Now we shift the origin from the centre along the x or y or z axis by
$100 h^{-1} \, {\rm Mpc}$ in the redshift space distribution of the
simulated galaxies. The resulting anisotropies as a function of $r$,
$\theta$ and $\phi$ are shown in the top right, middle right and
bottom right panels of \autoref{fig:millrsd} respectively. We see in
the top right panel of \autoref{fig:millrsd} that the observed
anisotropies change at smaller radii when the origin is shifted along
x, y or z directions than when it is not shifted. This tells us that
the distribution is anisotropic under such shifts. In the middle right
and bottom right panels we show the anisotropies in analogous
situations but as functions of $\theta$ and $\phi$ respectively. The
results in these panels show that the resulting anisotropies under
such shifts and without shift appears to be similar when measured as
functions of $\theta$ and $\phi$ whereas the level of anisotropies are
expected to be different in this case when the origin is shifted from
the centre. The top right panel of \autoref{fig:millrsd} agrees quite
well with this expected behaviour but the middle right and bottom
right panels do not exhibit these differences. The redshift space
distortions are caused by the radial component of peculiar velocities
which distorts the structures along the line of sight. The method
presented here may not be able to capture the anisotropies imprinted
in the details of distortions when looked in the polar and azimuthal
directions as we are currently using only the number counts within the
solid angle bins which radially extends upto $r_{max}$. Using the
correlation functions instead of number counts may prove to be a
better bet here.

\section{CONCLUSIONS}

We present an information theory based method for testing isotropy in
a three dimensional distribution and test the method on some Monte
Carlo simulations of isotropic and anisotropic distributions. We find
that our method can effectively identify and characterize various
types of anisotropies and distinguish between them. We insert pockets
of different densities inside homogeneous and isotropic distributions
and find that the proposed method can effectively quantify the degree
of the resulting anisotropy and also determine the geometry of the
pockets introduced. We also consider spherically symmetric radially
inhomogeneous distributions and find that such anisotropy can be
easily characterized by our method. We then study the anisotropies
induced by the redshift space distortions by using a semi analytic
galaxy catalogue from the Millennium simulation and find that the
method can separate such anisotropies from a general one.  But in
general the distributions could be much more complex specially when
the observed anisotropy results from different possible combinations
of various types of anisotropies. Disentangling such anisotropies is
no doubt would be quite challenging. However our method could serve
the purpose of detecting anisotropies quite well. In future we plan to
analyze data from the Two Micron All Sky Survey (2MASS) and the Sloan
Digital Sky Survey (SDSS) to test the assumption of isotropy in the
present Universe.

One may also extend the present method in Fourier space. In this case
one requires to estimate the PDF of the Fourier mode amplitudes in
different volume elements and then apply the anisotropy measures
presented here. Alternatively one can compute the Kullback-Leibler
divergence of the PDF of the Fourier mode amplitudes across the
various volume elements to identify and assess any power
asymmetry. One may also quantify the phase entropy by defining an
information entropy on the set of Fourier phases in different
hemispheres and then compare the information content across different
hemispheres. The degree of non-uniformity in the level of phase
information could be used as a measure of anisotropy in this case.

An important caveat in the present method arises from the tiling
strategy adopted here. The $\theta-\phi$ scheme that we have
implemented here ensure identical sizes for all the volume elements but
they do not have identical shapes which make the prediction of
anisotropy difficult in general. A HEALPix tessellation
\citep{gorski1,gorski2} would be more appropriate and useful while
analyzing observations. We plan to incorporate the HEALPix scheme into
our method in its future applications to galaxy surveys.

The proposed method can be also applied in many problems in Cosmology
which requires tests of isotropy. For example it can be used to
investigate the issues like the anisotropic distribution of galactic
satellites \citep{zentner} and anisotropic distribution of subhalos
inside dark matter halos and the cosmic web \citep{kang}.  The method
can be also used further to test for any hemispherical asymmetry in
the angular distribution of galaxy clusters \citep{bengaly1} and
gamma-ray bursts \citep{briggs}. Finally we note that the method
presented here has the desired ability to identify and characterize
any signals of anisotropy present in a distribution and it can be also
suitably adapted for different types of datasets from other
cosmological observations to efficiently explore the issue of Cosmic
isotropy.

\section{ACKNOWLEDGEMENT}
I sincerely thank an anonymous referee for a thorough review with
constructive suggestions which significantly helped to improve the
draft. The author would like to acknowledge CTS, IIT Kharagpur for the
use of its facilities for the present work. The author would also like
to acknowledge IUCAA, Pune for providing support through the
Associateship Programme.

The Millennium Simulation data bases \citep{lemson} used in this paper
and the web application providing online access to them were
constructed as part of the activities of the German Astrophysical
Virtual Observatory.

\bsp	
\label{lastpage}
\end{document}